\documentclass{article}
\usepackage{arxiv}
\usepackage[utf8]{inputenc} 
\usepackage[T1]{fontenc}    
\usepackage{url}            
\usepackage{booktabs}       
\usepackage{nicefrac}       
\usepackage{microtype}      
\usepackage{lipsum}		
\usepackage{graphicx}
\usepackage{natbib}
\usepackage{doi}
\usepackage{dblfloatfix} 
\usepackage[utf8]{inputenc}
\usepackage{amsmath}
\usepackage{amsfonts}
\usepackage{amssymb}
\usepackage{verbatim} 
  
\usepackage{natbib}
\usepackage{enumitem}  
\setlist{nolistsep,leftmargin=0.5 cm}
\usepackage{color}
\makeatletter
\newcommand{\pushright}[1]{\ifmeasuring@#1\else\omit\hfill$\displaystyle#1$\fi\ignorespaces}
\newcommand{\pushleft}[1]{\ifmeasuring@#1\else\omit$\displaystyle#1$\hfill\fi\ignorespaces}
\makeatother
\usepackage{multirow} 
\usepackage{array}    
\newcolumntype{P}[1]{>{\centering\arraybackslash}p{#1}} 
\newcolumntype{M}[1]{>{\centering\arraybackslash}m{#1}} 
\usepackage[final]{showkeys} 
\usepackage[nopostdot,nonumberlist]{glossaries}
\makeglossaries
 \usepackage{placeins}  
\usepackage{authblk} 
\usepackage{csquotes}	
\usepackage{rotating} 
\usepackage{dsfont}

\title{On CFD Numerical Wave Tank Simulations: Static-Boundary Wave Absorption Enhancement Using a Geometrical Approach}

\renewcommand{\undertitle}{A Preprint}	

\date{}

\author[1,2]{Muhannad W. Gamaleldin\thanks{Corresponding Author. \newline muhannad.gamaleldin@unimelb.edu.au \newline muhannad.wg@alexu.edu.eg }}
\author[1,3]{Alexander V. Babanin}
\author[4,5]{Amin Chabchoub}
\affil[1]{Department of Infrastructure Engineering, The University of Melbourne, Victoria 3010, Australia.}
\affil[2]{Mechanical Engineering Department, Alexandria University, Alexandria 21544, Egypt.}
\affil[3]{Laboratory for Regional Oceanography and Numerical Modeling, Qingdao National Laboratory for Marine Science and Technology, Qingdao 266237, China.}
\affil[4]{Centre for Wind, Waves and Water, School of Civil Engineering, The University of Sydney, NSW 2006, Australia.}
\affil[5]{Marine Studies Institute, The University of Sydney, NSW 2006, Australia.}
\hypersetup{
pdftitle={On CFD Numerical Wave Tank Simulations: Static-Boundary Wave Absorption Enhancement Using a Geometrical Approach},
pdfsubject={Ocean engineering, CFD},
pdfauthor={Muhannad W.~Gamaleldin, Alexander V.~Babanin, Amin Chabchoub},
pdfkeywords={Numerical Wave tank, static boundary absorption, CFD, Wave-structure interaction},
}

\begin{document}
\maketitle

\begin{abstract}
The present study aims to extend the applicability of the static-boundary absorption method in phase-resolving CFD simulations outside the conventional shallow-water waves limit. Even though this method was originally formulated for shallow-water waves based on the conventional piston type wavemaker, extending its use to deeper water conditions provides a more practical and computationally cost efficient solution compared to other available numerical wave absorption alternatives. For this sake, absorption of unidirectional monochromatic waves in a semi-infinite flume by means of a static wall is investigated theoretically and numerically. Moreover, implementation to a practical wave-structure interaction application is investigated numerically and experimentally. A phase-resolving numerical model based on the Reynold-averaged Navier-Stokes (RANS) equations is implemented using the open source C\texttt{++} toolbox OpenFOAM\textsuperscript{\textregistered}. The study presents the performance of the static-boundary method, in a dimensionless manner, by limiting the depth at which the active-absorption conventional-piston velocity profile is introduced; as a function of incident wave conditions. Moreover, it is shown that the performance of the static-boundary method can be significantly enhanced where wave reflection was reduced to about half of that of the conventional setup in deep-water conditions. Furthermore, the absorption depth is correlated to the incident wave conditions; providing an optimization framework for the selection of the proper dimensions of an absorbing wall. Finally, wave-structure interaction experimental tests were conducted to validate the numerical model performance; which shows an acceptable agreement between the model and the experimental observations. The proposed limiter is straight forward to be applied in pre-existing wave-structure interaction CFD solvers, without the need of code modifications.\\
\end{abstract}

\section{Introduction}
With the depletion of the relatively easily accessed resources inland and nearshore, marine industries have been marching further into deeper waters. This, in turn, led to increasing attention of the scientific community toward ocean engineering applications taking place in deep-water conditions. Generally speaking, phase-resolving CFD numerical simulations of ocean engineering applications are inherently expensive for a multitude of reasons such as interface tracking methods and the wide scale of physical processes that need to be resolved. However, one major challenge is the spatial boundedness of a numerical domain, in which waves need to be artificially absorbed at the domain bounds to mimic their corresponding real-life situations. In other words, wave reflection off numerical domain bounds ought to be prevented to avoid adversely affecting tested subjects. Consequently, a number of wave absorption techniques have been devised in literature; which can be classified into \emph{internal}, \emph{dynamic-boundary} and \emph{static-boundary} methods \citep{review}. First for the internal methods, also referred to as \emph{passive} methods, the wave motion is absorbed from the computational domain by dedicated zones inside the domain rather than using its boundaries; hence the name \emph{internal}. This can be done, for instance, by geometrically \citep{beach01,beach02} or numerically \citep{beach03,beach05} modifying an absorbing zone to imitate a \emph{dissipative beach} or \emph{sponge layers}. Another common instantiation is the use of a \emph{relaxation zone} \citep{relaxationZone4,waves2Foam,relaxationZone5} where the numerical solution is gradually blended to a desired wave motion and free surface profile. Overall, internal absorption methods are relatively simple and straight-forward to apply, compared to other methods. However, they increase the numerical expense due to their need to allocate relatively large domain zones for optimum wave absorption with minimal reflection; specially for longer waves. For instance, in \citep{exampleFreakWave,exampleAmini} it was needed to allocate a three-wavelength long zone to produce stable absorption.

\par Second, the dynamic-boundary methods where waves are absorbed using the numerical domain boundaries \citep{dynamicAWA,numericalAWA}; assimilating experimental \emph{active wave absorption} systems \citep{Milgram}. This is done by monitoring incident waves to the domain's absorbing termination/wall and moving it correspondingly, hence the name \emph{active}, in such a way complying with the wave motion and preventing reflection off the wall \citep{reviewAWA}. This in turn gives rise to the need of using computational techniques to simulate the dynamic walls \citep{reviewDynamicWalls}, posing a considerable numerical toll to the problem's overall computational cost. 

\par Finally, the static-boundary absorption method, which is the scope of the present study, where the concept is very close to the previous type except that the boundary is stagnant. This is simply because, in the numerical realm, there is no need to physically move a wall to generate or absorb waves; instead, motion of a wave absorbing wall is \emph{modeled} by artificially imposing the desired velocity profiles on the stationary boundary as a Dirichlet boundary condition. Consequently, this method is distinguished being the one with the least computational cost amongst other available methods and a practical choice in numerically demanding simulations \citep{platform2}. However, this absorption method suffers a severe limitation being applicable only to shallow water conditions due to the use of the conventional piston wavemaker in its formulation, which assumes a uniform velocity profile over the water depth. Clearly, this leads to restricted performance in deeper water conditions, majorly, due to the mismatch between the shallow and deeper water wave particle kinematics. For instance in \citep{pablo1}, it was reported that $11.2\%$ of wave amplitude was reflected when this method was used just about $13\%$ inside the intermediate-water condition range. Some alleviations have been proposed such as the combination with mesh stretching or other passive method \citep{review}, free surface imposition \citep{torres}, or even completely neglecting the piston profile and replacing it with an analytical expression based on a water wave theory \citep{pabloX}. For more detailed reviews about available techniques, readers are referred to \citep{review,beach04,relaxationZone6}.
\par In the present study, however, revisiting the classical wavemaker theory, one can see that a wavemaker with a reduced displacement toward the bottom, such as the flap wavemaker, obviously has better performance in deeper water conditions compared to the conventional piston. This is because the flap wavemaker kinematic velocity profile better matches the deep water particle kinematic velocity profile, compared to the uniform velocity profile posed by the piston wavemaker \citep{Dean}. Interestingly, this remark, however, has not been investigated in the aforementioned implementations of active wave absorption techniques in CFD numerical simulations. The present study aims to extend the range of applicability of the static-boundary absorption method beyond the conventional shallow-water limit. In what follows, the classical problem of wave absorption by a wall is revisited and investigated from a hydrodynamical perspective. This is done by modifying the conventional piston with a \emph{limited absorption depth} $\alpha h$ to better match the wave kinematics in deeper water conditions. Moreover, this depth is correlated to the incident wave conditions, providing an optimization framework for the selection of the proper piston dimensions. Additionally, implementation in two and three dimensional phase-resolving CFD numerical models is presented and validated against theory and physical experiments; respectively. It is worth highlighting that the proposed limiter is easy and straight forward to be applied in pre-existing numerical models/packages without any code modifications. This work is an in-depth analysis and benchmarking for the author's presented work at \citep{mw1,mw3,mw5}.\\

\section{The wavemaker theory}
\begin{figure}[!t] 
\centering
\includegraphics[width=0.55\textwidth]{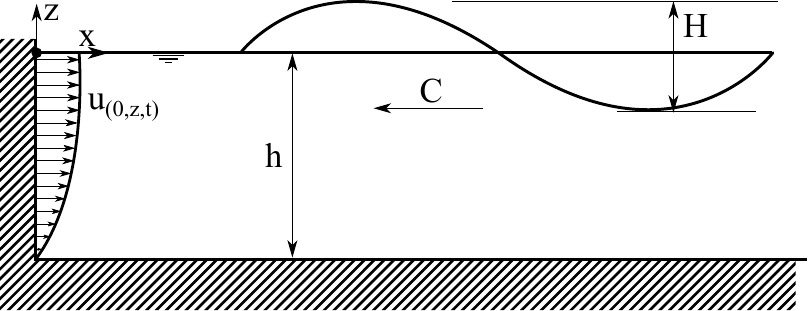}
\caption{Schematic diagram of the interaction of a Dirichlet velocity profile with an incident wave train.}
\label{fig:problemSetup}
\end{figure}

In what follows, the interaction of a reciprocating wall with a \emph{linear} unidirectional long-crested monochromatic wave is investigated analytically; using the potential flow theory. In a numerical context, this wall is merely a time-dependent velocity profile defined as a Dirichlet boundary condition. As shown in Fig. \ref{fig:problemSetup}, the problem at hand is to study the influence of the time-dependent velocity profile on an incident wave train propagating toward the wall at which velocity profile is introduced. If the frequency of the velocity profile is chosen to be equal to that of the incident wave, the problem at hand can be related to the concept of \emph{mechanical wave-absorbers}, also known as \enquote{active water-wave absorbers} by \citep{Milgram}; where energy of an incident wave train may be absorbed by a mechanical paddle. The underlying concept is explained by the following steps. First, the horizontal velocity profile $u(0,z,t)$ at the wall can be re-expressed as follows:
\begin{gather}
u(0,z,t) = u(z) \cos(\omega t)
\end{gather}
Where $\omega$ is the frequency of oscillation of the introduced velocity profile. After that, since the velocity profile is periodic, it might be replaced by moving the wall instead to generate the same velocity profile as follows:
\begin{gather}
u(z) \cos(\omega t) = \frac{S(z)}{2} \omega \cos(\omega t) \label{eqn:KBC00}
\end{gather}
Where $S(z)$ is the corresponding stroke profile of the reciprocating wall. Therefore, if the wall moves in such a way it becomes invisible to the incident wave; the wave energy is completely absorbed by the wall with no reflections. In other words, the wall is moved by the incident waves and complies to their motion \citep{Milgram}. Thereby, what happens here is, more or less, the opposite of the classical mechanical wavemaker problem; in which the motion of the wall causes the wave \citep{Dean,Ursell}. As a result, in the present context, the interaction of different subsurface profiles with incident waves is modelled as a mechanical wavemaker problem; linked together by the kinematic boundary condition Eq. (\ref{eqn:KBC00}).

\begin{figure}[!b] 
\centering
\includegraphics[width=0.55\textwidth]{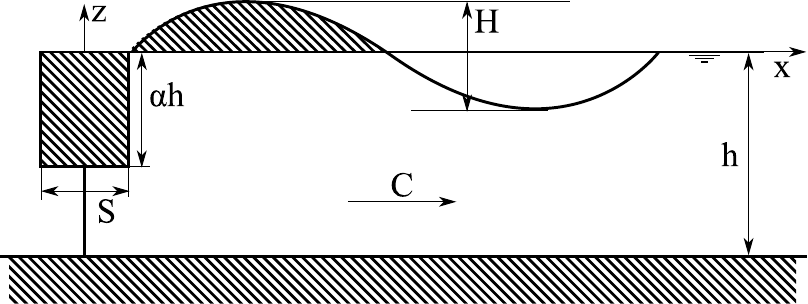}
\caption{Schematic diagram of the proposed step wavemaker.}
\label{fig:stepWaveMaker}
\end{figure} 

\begin{figure}[!t] 
\centering
\includegraphics[width=0.85\textwidth]{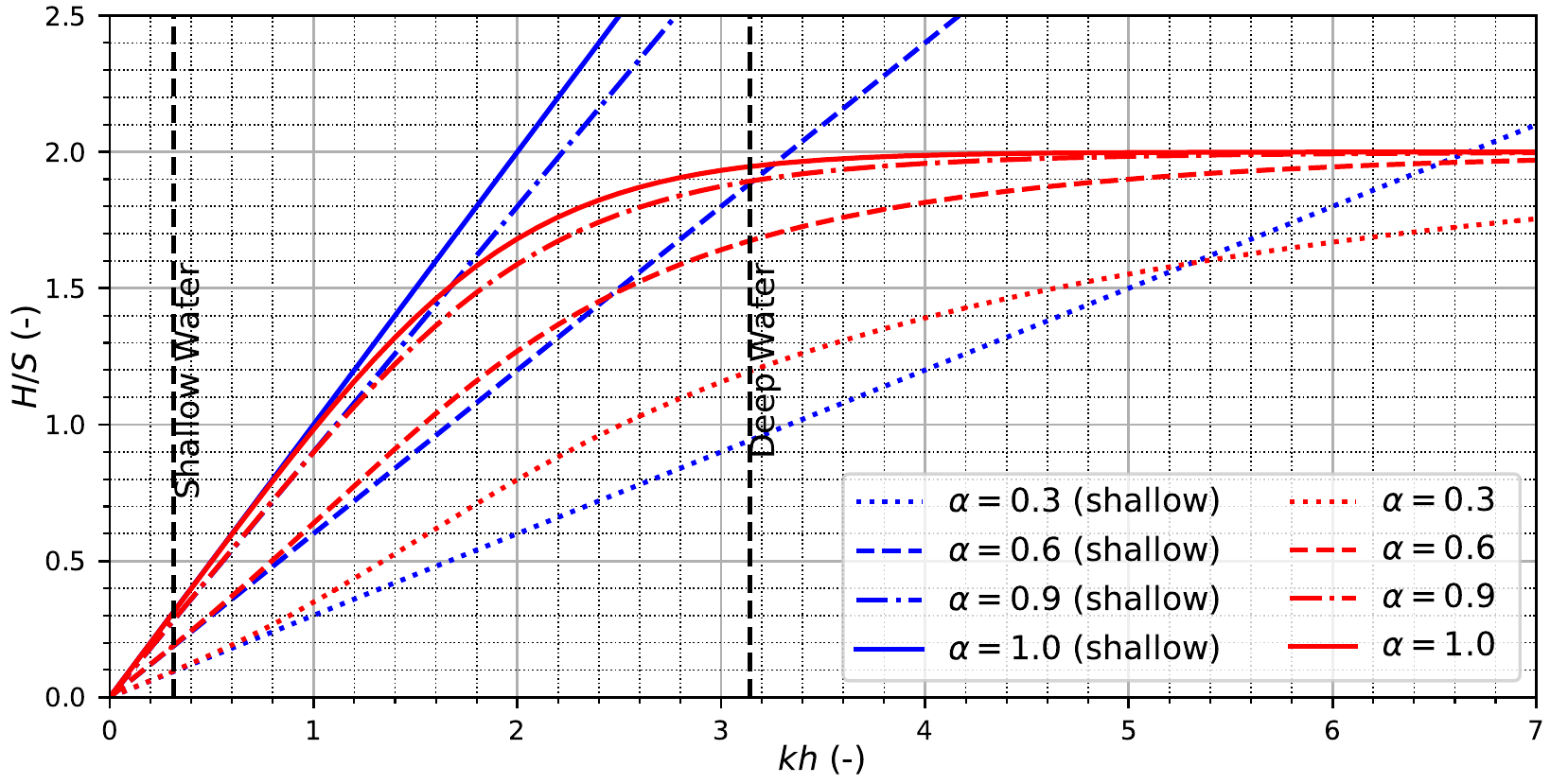} 
\caption{Wave-height-to-stroke ratio ($H/S$) versus relative depth ($kh$) for various values of $\alpha$. The label \enquote{shallow} indicates the simplified theory.}
\label{fig:allWaveMakersHS}
\end{figure} 

\par The proposed wavemaker design is the \emph{step wavemaker}, shown in Fig. \ref{fig:stepWaveMaker}. In comparison to the conventional piston wavemaker, the underlying idea is to examine the influence of introducing the velocity profile at a portion of the water depth ( limited absorption depth $\alpha h$) to better match the deep-water waves particle velocity profile and, as a result, reduce reflections. In what follows, the wave-height-to-stroke ratio is derived using the simplified and full wavemaker theories.

	\subsection{Simplified theory}
    In the present section, the relation between a mechanical wavemaker and resulted waves is derived based on a simplified assumption. The underlying concept is to assume that displaced water by a single stroke of a wavemaker is equal to accumulated water in the resulted wave crest; which is shown to be a valid approximation in the shallow-water conditions \citep{Galvin}. The relation between the wavemaker stroke and the resulted wave height ($H/S$) may be expressed by equating the displaced water by a full stroke to the volume under the wave crest as shown in Fig. \ref{fig:stepWaveMaker}, as follows:
      \begin{gather}
      S \alpha h = \int^{\lambda/2}_{0} \frac{H}{2} \sin(kx) dx\\
      S \alpha h = \frac{H}{k} \\
      \therefore \frac{H}{S} \Big|_{\text{step}} = \alpha k h \label{eqn:stepWaveMakerShallow}
      \end{gather}
      
      Where the expression Eq. (\ref{eqn:stepWaveMakerShallow}), plotted in Fig. \ref{fig:allWaveMakersHS}, is valid for relative depths of the range $k h < \pi / 10$; i.e. the shallow-water conditions. Clearly, when $\alpha = 1$, the step wavemaker reduces to Eq. (\ref{eqn:pistonWaveMakerShallow}); the simplified conventional piston wavemaker by \citep{Galvin}.
	\begin{gather}
    \frac{H}{S} \Big|_{\text{piston}} = k h \label{eqn:pistonWaveMakerShallow}
    \end{gather}

	\subsection{Complete theory}
	Assuming that generated waves are small-amplitude, long-crested and propagating in a potential, incompressible continuum; the governing equation for the velocity potential is expressed as follows:
   \begin{gather}
   \frac{\partial^{2}\phi}{\partial x^{2}} + \frac{\partial^{2}\phi}{\partial y^{2}}  = 0
   \end{gather} 
   Which is the Laplace equation boundary value problem. The boundary conditions are:
   \begin{align}
   \eta = \frac{1}{g} \frac{\partial \phi}{\partial t} \rightarrow & \text{ at }z=0 \label{eqn:DBC}\\
   -\frac{\partial \phi}{\partial z} = \frac{\partial \eta}{\partial t} \rightarrow & \text{ at }z=0 \label{eqn:KBC}\\
   -\frac{\partial \phi}{\partial z} = 0  \rightarrow & \text{ at }z=-h \label{eqn:BBC}
   \end{align}
   Where equations (\ref{eqn:DBC}), (\ref{eqn:KBC}) and (\ref{eqn:BBC}) are the dynamic and kinematic boundary conditions at the free surface and the impermeable sea bed condition; respectively. Moreover, the waves considered are propagating in the positive $x$-direction with no reflection from that end; as shown in Fig. \ref{fig:stepWaveMaker}. 
   
	Now, what is left is to satisfy the kinematic boundary condition on the wavemaker wall. To address this, first, the horizontal displacement of a wavemaker of a stroke profile $S(z)$ is described as follows:
   \begin{gather}
   x = \frac{S(z)}{2} \sin(\omega t)
   \end{gather}
	Therefore, the wavemaker wall can be described by the following surface function $F(x,z,t)$:
   \begin{gather}
   F(x,z,t) = x - \frac{S(z)}{2} \sin(\omega t) = 0
   \end{gather}
	Then, the kinematic boundary condition on the wavemaker wall is expressed as follows:
   \begin{gather}
   U.\hat{n} = \frac{- \partial F / \partial t}{|\nabla F|}  \\
   u-w \frac{\sin(\omega t)}{2} \frac{d S(z)}{dz} = \frac{S(z)}{2} \omega \cos(\omega t) \\
	u = \frac{S(z)}{2} \omega \cos(\omega t) \rightarrow  \text{ at  } x=\frac{S(z)}{2} \sin(\omega t)
   \end{gather}

	Finally, if the wavemaker displacement is considered to be relatively small, the kinematic boundary condition at the wavemaker can be linearized to be:
   \begin{equation}
   u(0,z,t) = \frac{S(z)}{2} \omega \cos(\omega t) \rightarrow \text{ at  } x=0 \label{eqn:kinematicBC}
   \end{equation}
   
	Now, after defining the boundary conditions, the general form of Laplace equation's solution satisfying the upper and lower boundary conditions is expressed as follows \citep{Havelock}:
   \begin{equation}
   \phi = A_{p} \cosh(k_{p}(h+z)) \sin(k_{p} x - \omega t) + \sum_{n=1}^{\infty} C_{m} e^{-k_{s(n)} x} \cos(k_{s(n)} (h+z)) \cos(\omega t) \label{eqn:phi}
   \end{equation}
   
	Where the subscripts $p$ and $s$ stand for \emph{progressive} and \emph{standing} waves generated by the wavemaker, respectively. Those standing waves, also referred to as \emph{evanescent} modes \citep{reviewAWA}, do not propagate and are rather locked at the vicinity of the wavemaker; decaying exponentially in the $x$-direction and are negligible two to three water depths of the wavemaker \citep{Ursell}. However, the coefficients $A_{p}$ and $C_{n}$ are found by applying the kinematic boundary condition at the wavemaker wall; and therefore, are dependent on the type of the wavemaker. Therefore, at $x=0$:
   \begin{align}
   u\big|_{\text{wavemaker}} &= u\big|_{\text{flow field}} \\
   \frac{S(z)}{2} \omega \cos(\omega t) &= -\frac{\partial \phi }{\partial x}
   \end{align}
   
	It can be shown that \citep{Dean}:
   \begin{align}
   A_{p} &= -I_{1}/I_{2} \text{ , where:} \label{eqn:Ap}\\    
   I_{1} &= \int^{0}_{-h} \frac{S(z)}{2} \omega \cosh(k_{p}(h+z)) dz \label{eqn:I1}\\
   I_{2} &= k_{p} \int^{0}_{-h}  \cosh^{2}(k_{p}(h+z)) dz \nonumber \\
         &= \frac{1}{4} \sinh(2 k_{p} h) + \frac{k_{p} h}{2}
   \end{align}
   
	Finally, the progressive wave height, relatively away from the wavemaker, can be linked to the wavemaker type as follows:
   \begin{gather}
   \eta = \frac{H}{2} \cos(k_{p} x - \omega t) \\
   \frac{1}{g} \frac{\partial \phi}{\partial t} \Big|_{z=0} = \frac{H}{2} \cos(k_{p} x - \omega t)\\
   \therefore H = \frac{-2A_{p}}{g} \omega \cosh(k_{p} h) \label{eqn:HoverSequation}
   \end{gather}   
   
	As such, by substituting an expression for $A_{p}$, which is dependent on the wavemaker shape, into Eq. (\ref{eqn:HoverSequation}); the \emph{wave-height-to-stroke} ratio ($H/S$) can be found. An expression $S(z)$ for the suggested step wavemaker, shown in Fig. \ref{fig:stepWaveMaker}, is proposed to be as follows:
      \begin{gather}
      S(z) = S ( \mathds{1}(z+\alpha h) - \mathds{1}(z) ) \label{eqn:stepS}
      \end{gather}
      
      Where $\mathds{1}(z)$ is the Heaviside unit-step function, i.e.:
      \begin{gather}
      \mathds{1}(z)
      \begin{cases}
      0                  & \text{for } z<0 \\
      1  & \text{for } z \geqslant	0
      \end{cases}
      \end{gather}
      
      Then, substituting into Eq. (\ref{eqn:I1}) we get:
      \begin{gather}
      I_{1} = \int_{-h}^{-\alpha h} 0 dz + \int_{-\alpha h}^{0} \frac{S}{2} \omega \cosh[k_{p} (h+z)] dz \\
      \therefore I_{1} = \frac{S \omega }{ 2 k_{p}} [\sinh(k_{p} h) - \sinh(k_{p} h (1-\alpha))]
      \end{gather}
      
      Now, substituting into Eq. (\ref{eqn:Ap}) we get:
      \begin{gather}
      A_{p} = \frac{- I_{1}}{I_{2}} = \frac{- S \omega [ \sinh(k_{p} h) - sinh (k_{p} h (1-\alpha))]}{2 k_{p} [\frac{1}{4} \sinh(2 k_{p} h) + \frac{k_{p} h}{2} ]}
      \end{gather}
      
      Finally, the wave-height-to-stroke ratio is obtained by substitution into the free surface expression, Eq. (\ref{eqn:HoverSequation}), as follows:
      \begin{gather}
      \frac{H}{2} = \frac{- A_{p}}{g} \omega \cosh(k_{p} h )\\
       \frac{H}{S} \Big|_{\text{step}} = \frac{4 \sinh(kh) [\sinh(kh) - \sinh(kh (1-\alpha)) ] }{ \sinh(2kh) + 2kh } \label{eqn:stepWaveMakerDeep}      
      \end{gather} 
      
      Equation (\ref{eqn:stepWaveMakerDeep}) is plotted in Fig. \ref{fig:allWaveMakersHS}, for several values of $\alpha$. Again, we can see in the figure that when $\alpha=1$, the step wavemaker reduces to the conventional piston wavemaker; described by the following expression \citep{Dean}:
	\begin{gather}
	\frac{H}{S} \Big|_{\text{piston}} = \frac{2 (\cosh(2k_p h) - 1 )}{\sinh(2 k_p h) + 2 k_p h} \label{eqn:pistonWaveMakerDeep}
	\end{gather}

	\subsection{Static-wall absorption}
   In what follows, implementation of the previous analysis as a Dirichlet boundary condition on the wave absorbing wall is presented. Going back to the original form of the problem, shown earlier in Fig. \ref{fig:problemSetup}, the step wavemaker can be transformed into a subsurface velocity profile $u(0,z,t)$ by substituting the proposed step wavemaker profile Eq. (\ref{eqn:stepS}) into the kinematic boundary condition Eq. (\ref{eqn:kinematicBC}) as follows:
   \begin{gather}
   u(0,z,t) = \frac{S}{2} ( \mathds{1}(z+\alpha h) - \mathds{1}(z) ) \omega \cos(\omega t) \label{eqn:new1}
   \end{gather}
   
   Where $S$ here might be thought of as an amplitude which is dependent on the incident wave height $H$ and the relative water depth $kh$.\\
   Thereby, the instantaneous velocity profile of the subsurface velocity profile can be now linked to the instantaneous free surface elevation $\eta(0,t)$, measured at the wall, by substituting the $H/S$ expression (\ref{eqn:stepWaveMakerDeep}) into Eq. (\ref{eqn:new1}) we get:
   \begin{equation}
    u(z,t)  = \omega \frac{H}{2} \cos(kx-\omega t) ( \mathds{1}(z+\alpha h) - \mathds{1}(z) ) \frac{ \sinh(2kh) + 2kh }{4 \sinh(kh) [\sinh(kh) - \sinh(kh (1-\alpha)) ] }
   \end{equation}
   Therefore:
   \begin{equation}
    u(z,t)  = \omega ( \mathds{1}(z+\alpha h) - \mathds{1}(z) ) \eta(0,t) \frac{ \sinh(2kh) + 2kh }{4 \sinh(kh) [\sinh(kh) - \sinh(kh (1-\alpha)) ] } \label{eqn:stepCurrents}
   \end{equation}
   
   Where, Eq. (\ref{eqn:stepCurrents}) is the corresponding subsurface instantaneous velocity profile of the step profile. As seen in the equation, once the free surface elevation is monitored, the velocity profile needed to absorb the incident wave is defined. According to the potential flow theory, if this profile is applied at a wall, the incident wave will be absorbed; with no reflected components. Practically speaking, reflected components will be present and need to be corrected for; which is discussed in the following sections.
   
   \par Last  but not least, it is worth mentioning here that a simplified version of the previous procedure is what is being used in the C\texttt{++} library OpenFOAM; specifically, the \texttt{IHFOAM} and \texttt{OlaFOAM/OlaFlow} packages developed by \citep{pablo1}. This is done by substituting the shallow-water approximation of H/S for the conventional piston wavemaker Eq. (\ref{eqn:pistonWaveMakerShallow}), into the kinematic boundary condition Eq. (\ref{eqn:kinematicBC}) as follows:
   \begin{gather}
   u(0,z,t) = \frac{\omega}{2}  \frac{H}{kh} \cos(\omega t) \\
   u(0,z,t) = u(0,t) = \frac{C}{h} \text{ } \eta(0,t) \label{eqn:IhOlaFoamVersion}
   \end{gather}
   
   Where, $C$ is the incident wave celerity; calculated using the shallow-water approximation $C = \sqrt{g h}$. Moreover, it could have also been calculated using the dispersion relation. In an experimental context, however, it is also worth highlighting here that monitoring the celerity of incident waves is a challenging aspect, specially outside the shallow-water conditions and dispersive waves, and a number of other techniques have been used in the context of using discrete-time systems and digital filters \citep{digitalFilters01,digitalFilters02,dispersiveWaves}; which is outside the scope of the present study which investigates the problem within a hydrodynamic scope.\\

\section{Numerical model}
The numerical package of choice in the present study is the \emph{Open-source Field Operations And Manipulations} (OpenFOAM). Unlike commercial codes, OpenFOAM is not the type of a black box where the user can modify any step of the solution process by modifying the source code and it covers a wide range of academic and industrial problems (liquid sprays, external flows, multiphase flows...etc.). OpenFOAM is a C\texttt{++} library utilizing \emph{object-oriented-programming} to present, discretize and solve computational partial differential equations (PDEs) \citep{openfoam,openfoam03}. OpenFoam CFD solvers proved strong potentials and capabilities by numerous case studies in the literature. For instance, in the present scope, correctly modelling of up to 8\textsuperscript{th}-order harmonics for regular waves over a submerged bar by \citep{validation06} and \citep{validation07}. Moreover, \citep{beach05} evaluated the performance of OpenFOAM, specifically the InterFOAM module, to simulate interactions of non-linear waves with offshore structures; showing good agreement with experimental observations. In what follows, details of the numerical model implemented in the present study is presented; with this \texttt{font} used as a semantic markup for OpenFOAM specific verbatim entries and expressions.\\
\begin{table}[!t]  
\centering
\caption{Finite volume discretization schemes listed in OpenFOAM's \texttt{fvSchemes} dictionary.}
\label{tab:schemes}
\begin{tabular}{ll}
\hline
\multicolumn{1}{c}{Transport Term}     & \multicolumn{1}{c}{Discretization Scheme}\\ \hline
\multirow{3}{*}{\texttt{divSchemes}}   & \texttt{div(phirb,alpha) $\rightarrow$ Gauss interfaceCompression}     \\
                                       & \texttt{div(rhoPhi,U) $\rightarrow$ Gauss limitedLinearV 1} \\ 
                                       & \texttt{div(phi,alpha) $\rightarrow$ Gauss vanLeer} \\
\texttt{gradSchemes}                   & \texttt{Gauss linear}               \\
\texttt{snGradSchemes}                 & \texttt{corrected}                  \\
\texttt{interpolationSchemes}          & \texttt{linear}                     \\
\texttt{timeScheme}                    & \texttt{Euler}                      \\ \hline
\end{tabular} 
\end{table}  
\begin{table}[!b]  
   \centering
   \caption{Computational domain boundary conditions in OpenFOAM's conventions.}
   \label{tab:bounCon}
   \resizebox{\textwidth}{!}{%
   \begin{tabular}{ccccc}
   \hline
   \multirow{2}{*}{Boundary} &       & \multicolumn{3}{c}{Boundary Condition} \\ \cline{2-5} 
                             & Field & $p$          & $U$         & $\gamma $         \\ \hline
   Inlet wall               & & \texttt{fixedFluxPressure}         & \texttt{waveVelocity}         & \texttt{waveAlpha}             \\
   top wall                 & & \texttt{totalPressure}          & \texttt{pressureInletOutletVelocity}         & \texttt{inletOutlet}             \\
   Bottom wall              & & \texttt{fixedFluxPressure}          & \texttt{fixedValue}         & \texttt{zeroGradient}             \\
   Outlet wall $\alpha h$   & & \texttt{fixedFluxPressure}          & \texttt{waveAbsorption2DVelocity} (Eq. (\ref{eqn:IhOlaFoamVersion}))         & \texttt{zeroGradient}             \\
   Outlet wall $(1-\alpha)h$& & \texttt{fixedFluxPressure}          & \texttt{fixedValue}         & \texttt{zeroGradient}             \\ \hline
   \end{tabular}
   }
   \end{table}  
   \subsection{Governing equations and discretization schemes}
   \label{sec:numGovEqn}
   The numerical technique implemented in the present study is classified as a \emph{CFD phase-resolving} model \citep{phaseResolvingModels}. This is because displacement of the air-water interface is resolved by means of a sufficiently fine computational grid in comparison with the wave length. Transport equations are expressed for two inviscid incompressible immiscible fluids of different phases: air and water. The governing transport equations are expressed as follows:
\begin{gather}
\nabla \cdot U = 0 \label{eqn:CE}\\
\frac{\partial (\rho U)}{\partial t} + \nabla \cdot (\rho U U) = - \nabla p + \rho f_{b} \label{eqn:ME} \\
\frac{\partial \gamma}{\partial t} + \nabla \cdot (U \gamma) = 0 \label{eqn:phase}
\end{gather}

	Equations (\ref{eqn:CE}), (\ref{eqn:ME}) and (\ref{eqn:phase}) are the mass conservation, momentum transport, and phase-fraction transport equations. Where, $U$ is the velocity vector at any point in the numerical domain, $\gamma$ is the phase-fraction indicator, $\rho$ is the fluid density, $p$ is the pressure and $f_{b}$ is the term that includes the effect of body forces per unit fluid mass; specifically, the gravity and surface-tension forces. The phase-fraction indicator is a piece-wise function that equals zero and one in the computational domain cells consisting of gas and liquid, respectively. Thereby, at the interface region between the two fluids, the phase-fraction indicator can take values between zero and one. Moreover, this change is, and ought to be, sharp to form a high phase-fraction gradient. Furthermore, this sudden change/gradient at the interface region needs to be numerically preserved and propagated to provide realistic simulations; resembling the coexistence of two immiscible fluids in the computational domain.

	\par The simulation procedure starts by initializing the computational domain by allocating each fluid type to its corresponding initial zone; hence, an interface is introduced. After that, the governing equations are solved to transport the fluid and preserve a sharp interface. A two-fluid model approach is implemented \citep{rusche} which is based on defining the contributions of each specific fluid to the velocity field $U$ on a phase-averaged basis, i.e.:
\begin{gather}
U = \gamma U_{liquid} + (1-\gamma) U_{gas}
\end{gather}

The phase-averaged velocity concept facilitates the ability to algebraically sharpen the interface regions, hence classified as algebraic-VOF, by means of the so called \enquote{compression} term \citep{rusche,berberovic}. The compression term is added to the right hand side of the phase-fraction equation Eq. (\ref{eqn:phase}) as follows:
\begin{gather}
\frac{\partial \gamma}{\partial t} + \nabla \cdot (U \gamma)  + \nabla \cdot [U_{r} \gamma (1 - \gamma)] = 0 \label{eqn:phaseWithCompression}
\end{gather}

	Where, $U_{r}$ is the relative velocity vector between the two phases (i.e. $= U_{liquid} - U_{gas}$). Even though, the new term in equation Eq. (\ref{eqn:phaseWithCompression}) vanishes analytically\footnote{The compression term $\nabla \cdot [U_{r} \gamma (1 - \gamma)]$ vanishes in continuum formulation as $\gamma$ equals either $0$ or $1$; the term exists only at the interface region in numerical formulation as $\gamma$ forms a gradient between $0$ and $1$.}, it has a fundamental numerical role by compressing the interface region into a couple-of-cells thickness \citep{berberovic}. This method is broadly used in two-phase simulations in general and in sea waves simulations in specific; solved using the \enquote{multidimensional universal limiters for explicit solution} (MULES) solution algorithm, guaranteeing boundedness of the solution \citep{deshpande}. The OpenFOAM numerical solver of choice is the \texttt{olaFlow} solver which is an updated version of the the older \texttt{IHFoam} package \citep{pablo1}; based on OpenFOAM's well known interfacial flow solver \texttt{interFoam} \citep{openfoam03}.
	
	\par The selection and understanding of the discretization methods implemented forms a fundamental role in the numerical simulation; specifically, it directly influences solution fields conservativeness, boundedness and Transportiveness \citep{versteeg}. Table \ref{tab:schemes} summarizes the discretization schemes of choice, in OpenFOAM's semantics, which are specified in the finite-volume dictionary \texttt{fvSchemes}. Computational grid cell-center values are interpolated to the cell's face-centers using the \texttt{linear} interpolation scheme, declared in the \texttt{interpolationSchemes} sub-dictionary, as follows:
	\begin{equation}
	\psi_f = 0.5 \left( \psi_c + \psi_d \right)
	\end{equation}	 
	
	Where $\psi$ indicates an arbitrary field variable; and the subscripts $f$, $c$ and $d$ indicates centers of face, central-node and downstream-node, respectively. Therefore, interpolating face values using a second-order central differencing scheme. After that, calculations of the gradient terms are declared in the \texttt{gradSchemes} to be \texttt{Gauss linear}; indicating that the standard Gaussian integration of the finite volume method is used to calculate the face-center values from the cell-center values, using the \texttt{linear} interpolation scheme described earlier. Then the \texttt{snGradSchemes} sub-dictionary is used to declare the evaluation method of gradient components in the direction normal to a cell's face from the aforementioned calculated cell-based gradients:
	\begin{equation}
	\nabla^{\perp}_{f}  \psi = \hat{n} \cdot (\nabla \psi)_{f}
	\end{equation}
	
	Here, $\hat{n}$ is the unit normal vector to the the cell's face. The \texttt{divSchemes} is where the divergence schemes are declared for the convective terms of the transport equations. 
	The entry \texttt{div(rhoPhi,U)} resembles $\nabla \cdot (\rho U U)$ term in Eq. (\ref{eqn:ME}); set to be \texttt{Gauss limitedLinearV 1}, where face values are interpolated using the convection-dependent scheme \texttt{limitedLinearV} belonging to the \emph{Total Variation Diminishing} (TVD) class \citep{tvd}. This implies that the interpolation of the face values is influenced by the flow direction, limiting toward the upwind scheme. This is chosen instead of a central differencing one to reflect the transportivness nature of the problem for the convection terms. Then the entry \texttt{div(phi, alpha)}   resembles $ \nabla \cdot (U \gamma) $ term in Eq. (\ref{eqn:phaseWithCompression}) where the convection dependent Van Leer discretization scheme is selected; which is a high resolution nonlinear Fromm-based scheme  \citep{vanLeer}--- unbounded second-order accurate in space. The last entry for divergence terms is 	\texttt{div(phirb, alpha)} resembling the fundamental compressions term $ \nabla \cdot [U_{r} \gamma (1 - \gamma)] $ of Eq. (\ref{eqn:phaseWithCompression}). Here, it is worth mentioning that the aforementioned MULES algorithm utilizes a delimiter $L_{M}$ which is equal to $1$ and $0$ at the interface-zone and elsewhere, respectively. This is done as a mean to switch between using the high resolution  scheme at the interface region and the straight-forward convection scheme elsewhere to conserve computational cost \citep{rusche,deshpande}. Finally, temporal derivatives are declared at the \texttt{ddtSchemes} subdictionary to be the implicit first-order Euler scheme:
	\begin{equation}
	\frac{ \partial \psi }{ \partial t} = \frac{\psi - \psi^{\circ}}{\Delta t}	
	\end{equation}
	
	Where the superscript $\circ$ indicates the previous time-level solution.\\

   \subsection{Geometry and boundary conditions}
   \label{sec:GeomAndBC}
\begin{figure}[!t] 
\centering
\includegraphics[width=0.55\textwidth]{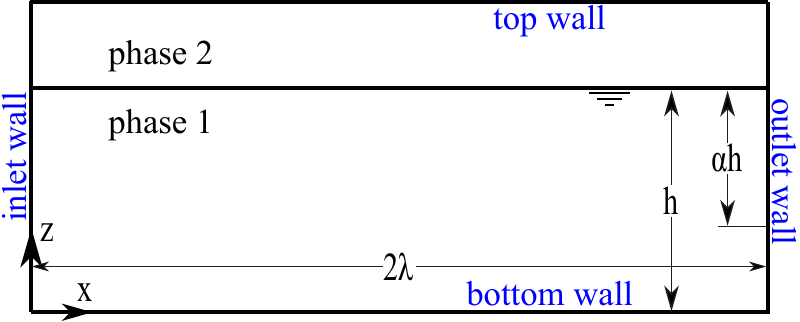}
\caption{Schematic diagram for the numerical domain setup.}
\label{fig:numDomain}
\end{figure}

	A computational domain is constructed, as shown in Fig. \ref{fig:numDomain}. The domain is $9~\operatorname{m}$, $3.8~\operatorname{m}$, and a unity in the x,z and y directions; respectively. The numerical domain is descretized into a structured multi-block mesh using the \texttt{blockMesh}\citep{blockMesh} utility in OpenFOAM. Three grid levels where created: coarse, medium and fine; with $0.8$, $1.4$ and $4$ million cells, respectively. This is done to ensure proper grid resolution where the number of cells per design wave height are $16$, $19$ and $31$ for the coarse, medium and fines grids, respectively. Moreover, to maintain a recommended value of grid refinement ratio $r \geq 1.3$ for the sake of the numerical verification calculations presented later in the succeeding section.
	
\begin{figure}[!t] 
   \centering
   \includegraphics[width=0.85\textwidth]{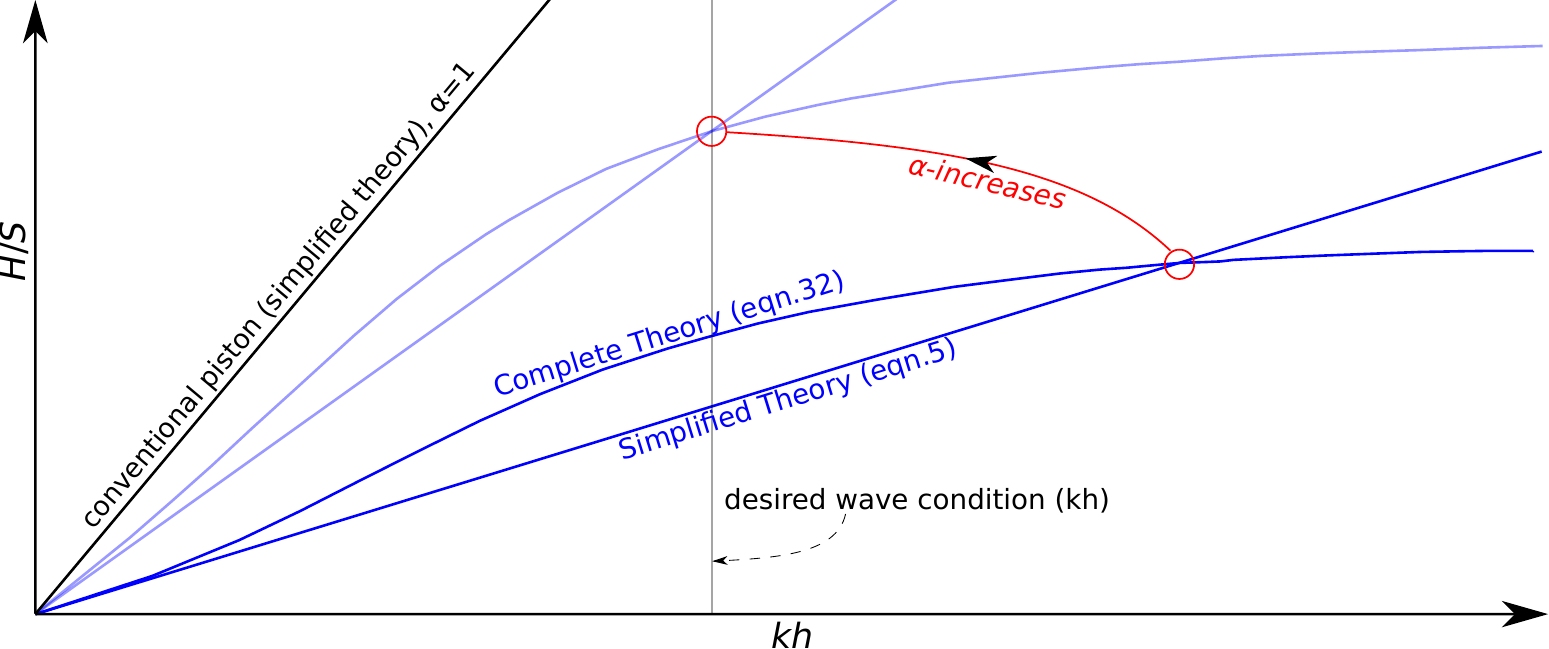} 
   \caption{A graphical representation of the underlying concept of step wavemaker tuning to meet a desired performance point.}
   \label{fig:pump}
\end{figure} 

	\par For the temporal resolution, adaptive time step is used to limit the maximum Courant–Friedrichs–Lewy (CFL) number in the numerical domain to $0.15$. Moreover, since the model used in the present study is a phase resolving one, the maximum CFL number of the air-water interface is also limited to a maximum value of $0.15$. Finally, the simulations where conducted to $60$ seconds of waves propagation.
	\par Table \ref{tab:bounCon} summarizes the boundary conditions of choice; in OpenFOAM's conventions. Since the theoretical analysis in the previous section tackles linear monochromatic waves, the simulated waves are selected to be so. Waves are generated by the \emph{inlet wall} using the Stokes 1\textsuperscript{st}-order theory \citep{Dean}. The generated waves at the inlet wall are $0.01~\operatorname{m}$ high, $1.7~\operatorname{s}$ period, $4.5~\operatorname{m}$ long and $3~\operatorname{m}$ deep. After that, waves are absorbed by the \emph{outlet wall} where the subsurface velocity profile is introduced along the $\alpha h$ portion of the entire water depth $h$; using the \texttt{OlaFlow} wave generation and absorption library. The \emph{top wall} and \emph{bottom wall} are set to be atmospheric and solid walls; respectively. Finally, \emph{phase 1} and \emph{phase 2} are set to be standard water and air; respectively. Simulations where conducted on the University of Melbourne's high performance computer Spartan \citep{spartan}.

\section{Results and discussion}
Hereafter, interaction of monochromatic waves with an active absorbing wall is investigated. Practically, incident waves will not be completely absorbed and some reflection will take place. This, in turn, leads to a partial standing wave field in the wave flume. As a result, the wave height will not be constant; instead, it varies sinusoidally along the channel. Consequently, wave reflection coefficient forms a fundamental relevant benchmarking quantity in the present study to evaluate performance of the wave absorbers investigated. The reflection coefficient can be found using a number of techniques such as using a carriage-mounted wave-height transducer or using the three gauges method by \citep{mansard}. Nevertheless, since tests in the present study is conducted in a numerical flume, the wave envelope is measured along the channel and the wave amplitude reflection coefficient $\epsilon_{r}$ is found as follows:
   \begin{gather}
   \epsilon_{r} = \frac{H_{r}}{H_{i}} = \frac{|\eta|_{\text{max}}-|\eta|_{\text{min}}}{|\eta|_{\text{max}} + |\eta|_{\text{min}}} \label{eqn:refCoe}
   \end{gather}  
   
   Where ${H_{r}}$ and ${H_{i}}$ are the reflected and incident wave heights, respectively. 

\begin{figure}[!b] 
\centering
\includegraphics[width=0.55\textwidth]{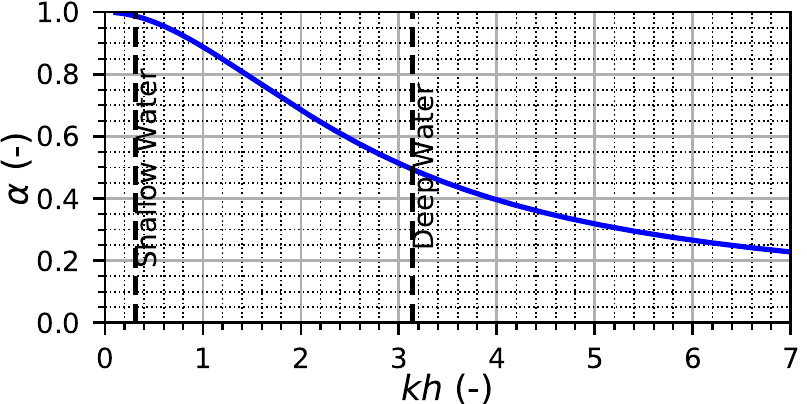} 
\caption{Dimensionless absorption depth ($\alpha$) versus relative depth ($kh$).}
\label{fig:optimumAlpha}
\end{figure} 
\begin{figure}[!t] 
\centering
\includegraphics[width=\textwidth]{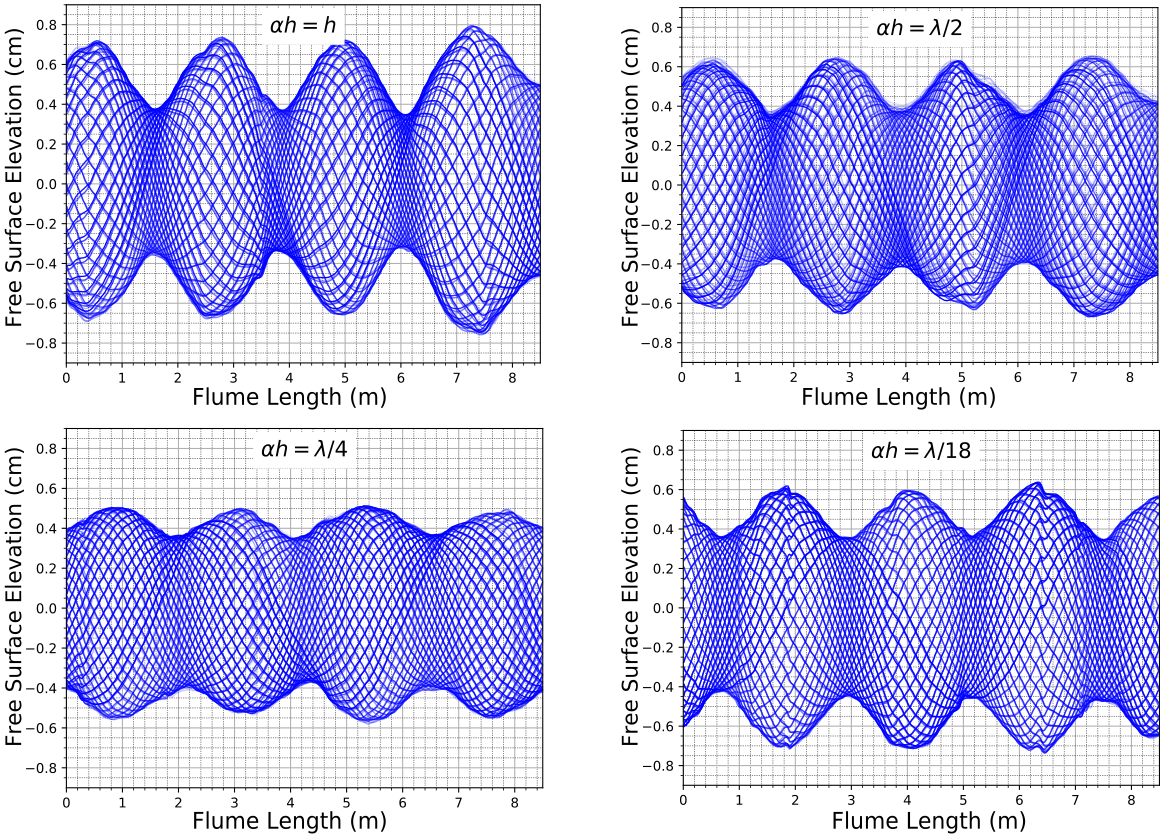} 
\caption{Temporal evolution of the water-air interface during a $10$ seconds duration for various limited absorption depths $\alpha h$.}
\label{fig:envelopes}
\end{figure} 
   \subsection{Active wave absorption enhancement}
	 As illustrated earlier in the theoretical modelling section, the subsurface velocity profile in the numerical package is calculated using Eq. (\ref{eqn:IhOlaFoamVersion}); which is valid only for shallow-water conditions because it assumes a uniform particle kinematic velocity profile over the entire depth. In other words, if a deep-water wave is considered, the value of $H/S$ calculated by the numerical model, based on the simplified wavemaker theory, will be far from the actual true value of $H/S$ of the complete wavemaker theory; as can be seen in Fig. \ref{fig:allWaveMakersHS}. So, the goal now is to correct the value of $H/S$ of the numerical model at deep-water conditions. In what follows a simple alleviation for this problem is proposed.\\
	\par A work around this problem is to set the numerical model to activate the velocity profile on a portion $\alpha h$ of the outlet wall instead of the whole depth $h$; assimilating the \emph{step wavemaker} modelled earlier by equations (\ref{eqn:stepWaveMakerShallow}) and (\ref{eqn:stepWaveMakerDeep}). The underlying idea is to make use of the step wavemaker versatility as the potential flow solution predicts; i.e. for a given wave condition $kh$, $\alpha$ can be varied to produce a desirable $H/S$ value. Moreover, the step wavemaker profile is expected to better match the velocity profile of a deep-water wave kinematic compared to the shallow-water uniform kinematic profile posed by the conventional piston type. This can be illustrated as follows:
		 \begin{gather}
		 \frac{H}{S} \Big|_{\text{Numerical}} = \frac{H}{S} \Big|_{\text{Exact}} \\
		 \therefore \frac{H}{S} \Big|_{ \text{Eq. (\ref{eqn:stepWaveMakerShallow})} } = \frac{H}{S} \Big|_{ \text{Eq. (\ref{eqn:stepWaveMakerDeep})} } \\
		 \alpha k h = \frac{4 \sinh(kh) [\sinh(kh) - \sinh(kh (1-\alpha)) ] }{ \sinh(2kh) + 2kh } \label{eqn:optimumAlpha}
		 \end{gather}
		 
   Therefore, by solving Eq. (\ref{eqn:optimumAlpha}), one can find the optimum value of $\alpha$ so that the value of $H/S$ calculated by the numerical model (using the shallow-water approximation Eq. (\ref{eqn:stepWaveMakerShallow})) comply with the corresponding $H/S$ given by the complete wavemaker theory Eq. (\ref{eqn:stepWaveMakerDeep}). Graphically, this can be conceived as tuning the performance curves (i.e. $kh$ versus $H/S$ curves) of a step wavemaker until a desired wave condition ($kh$) is met. Figure \ref{fig:pump} illustrates this concept where the value of $\alpha$ is varied in such a way that the interception point (i.e. solution of Eq. (\ref{eqn:optimumAlpha})) falls in the desired $kh$ range. Equation (\ref{eqn:optimumAlpha}) is deemed implicit and solved by trial and error as shown in Fig. \ref{fig:optimumAlpha}. Indeed, one can see in the figure that $\alpha$ is, more or less, a unity in the shallow-water range.\\
	\par For instance, for the wave conditions considered in the present study, the relative wave depth is $kh=4.18$. By substituting into Eq. (\ref{eqn:optimumAlpha}) we get $\alpha = 0.38$. This indicates that, for optimum wave absorption of the outlet wall with minimum reflection, the subsurface velocity profile is to be introduced along a depth $ \alpha h \approx \lambda/4 $. In what follows, a parametric case study is conducted to investigate the influence of varying the limited absorption depth $\alpha h$ on the performance of an active wave absorption wall in deep-water conditions.
\begin{table}[!t] 
\caption{Calculations of the GCI spatial discretization-error estimates.}
\begin{center}
 \begin{tabular}{ccc}
\hline
                     & $\varphi = \epsilon_r$ \scriptsize{(-)} & $ \varphi = A_{k}$ \scriptsize{(cm)} \\ \hline
$r_{21}$             & $1.69$                       & $1.69$          \\
$r_{32}$             & $1.3$                        & $1.3$           \\
$\varphi_{1}$        & $0.136$                      & $0.361$         \\
$\varphi_{2}$        & $0.127$                      & $0.368$         \\
$\varphi_{3}$        & $0.113$                      & $0.367$         \\
$P$                  & $2.85$                       & $2.87$          \\
$\varphi^{21}_{ext}$ & $0.139$                      & $0.359$         \\
$e^{21}_{a}$         & $7.22\%$                     & $2.05\%$        \\
$e^{21}_{ext}$       & $2.03\%$                     & $0.58\%$        \\
$GCI^{21}_{fine}$    & $2.6\%$                      & $0.73\%$        \\ \hline
\end{tabular} \\
\label{tab:GCIcalculations}
\end{center}
\end{table} 
	\subsection{Numerical model verification}
	Although analytical/theoretical solution is available to validate the numerical results against, estimating discretization errors is still relevant as numerical simulations are usually intended to simulate influences of physical events even when such validation data are absent. Moreover, it is also deemed important to argue whether the outcome of a numerical model is of a physical behaviour or, otherwise, random. In the present study, numerical uncertainty is investigated using the \emph{Grid Convergence Index} (GCI) method \citep{JFE,roache}; which is based on the \emph{Richardson Extrapolation} (RE) method \citep{re1,re2}. The GCI represents a measure of how much a computed value is distant off a numerical model's asymptotic value. 
	\par For this sake, the 	amplitude reflection coefficient ($\epsilon_{r}$) and the amplitude spectral peak ($A_{k}$) are chosen as the global variables of issue for the grid convergence test. Table \ref{tab:GCIcalculations} showcases the GCI calculations. It shows that the apparent spatial order of convergence of the numerical solution $P$ is found to be around $2.8$ for both variables. Moreover, values of GCI for the fine-grid solution for $\epsilon_{r}$ and $A_{k}$ are $2.6\%$ and $0.73\%$, respectively; indicating that the numerical results are relatively within acceptable margin of the asymptotic solution.
\begin{table}[!b]
\centering
\caption{Wave statistical quantities for various limited absorption depths $\alpha h$.}
\label{tab:reflection}
\begin{tabular}{ccccc}
\hline
$\alpha h$     & $h$    & $\lambda /2$ & $\lambda/4$ & $\lambda/18$ \\ \hline
$\epsilon_{r} (\%)$ & $28.1$ & $22.51$      & $12.66$     & $21.94$         \\
$A_{k} (cm)$        & $0.543$    & $0.484$          & $0.369$         & $0.54$         \\
$\sigma^{2} (cm^{2})$   & $0.178$& $0.141$      & $0.083$     & $0.173$         \\
$H_{s} (cm)$        & $1.69$    & $1.50$          & $1.15$         & $1.66$         \\ \hline
\end{tabular}%
\end{table}
	\subsection{Flow visualization and waves transmission}
	A parametric study is conducted to investigate the influence of varying the limited absorption depth $\alpha h$ on the performance of a wave absorbing wall. Four absorption depths $\alpha h$ are investigated: $h$, $\lambda/2 $, $\lambda/4$ and $\lambda/18$. Figure \ref{fig:envelopes} shows the temporal locus of the air-water interface during a $10$ second duration, where the interface is defined numerically by the set of cells where the \emph{phase fraction function} $\gamma = 0.5$.
	 \par For the first case $\alpha h = h$, this case resembles the conventional numerical setting where subsurface velocity profile is introduced along the whole depth. As have been addressed earlier in the theoretical analysis section, the velocity profile is introduced in such a way that a conventional piston wave absorber is used (i.e. equations (\ref{eqn:pistonWaveMakerShallow}) and (\ref{eqn:IhOlaFoamVersion})). And since this is based on the shallow-water conditions approximation, high reflection is anticipated from the absorbing wall. As seen in Fig. \ref{fig:envelopes}, reflection takes place and the wave height varies sinusoidally along the wave flume. Indeed, one can see that this is a manifestation of the classical \emph{partial standing wave} system. Bearing that in mind, the amplitude reflection coefficient $\epsilon_{r}$ can be calculated using Eq. (\ref{eqn:refCoe}) and is found to be $\epsilon_{r}=28.1\%$.
\begin{figure}[!t] 
   \centering
   \includegraphics[width=0.55\textwidth]{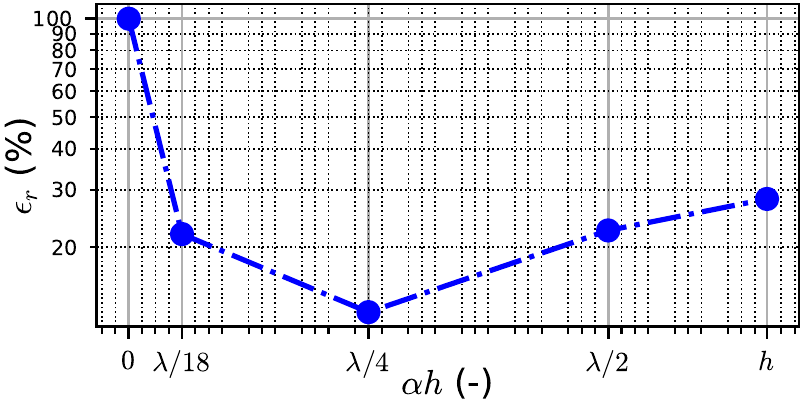} 
   \caption{Reflection coefficient ($\epsilon_{r}$) versus limited absorption depth ($\alpha h$).}
   \label{fig:reflectionCoe}
   \end{figure} 
   \begin{figure}[!b] 
   \centering
   \includegraphics[width=0.85\textwidth]{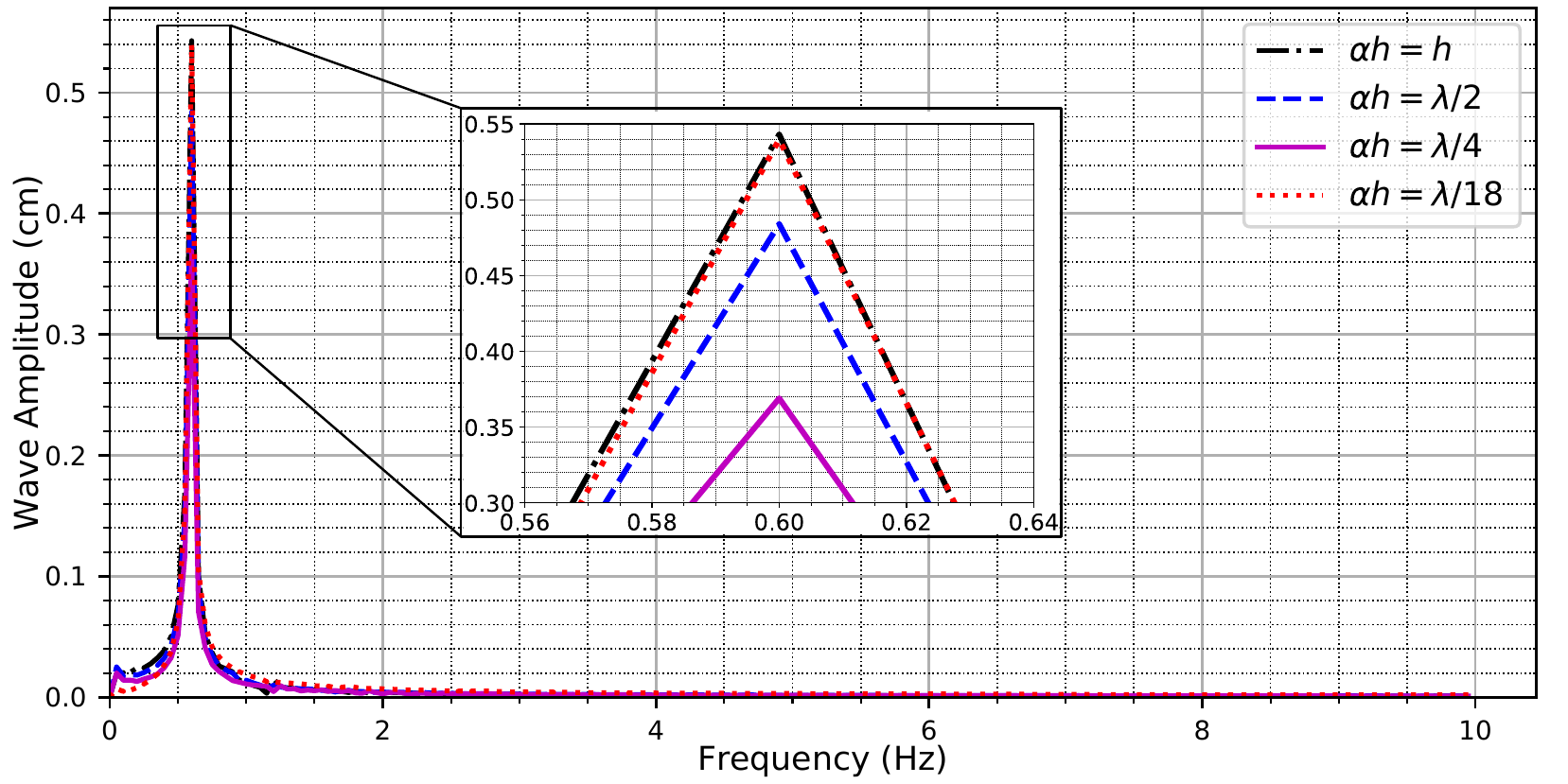} 
   \caption{Wave amplitude spectra measured by a wave gauge mounted at mid-flume; at a sampling frequency of $20$ Hz.}
   \label{fig:spectrum}
   \end{figure} 
   \begin{figure}[!t] 
      \centering
      \includegraphics[width=0.55\textwidth]{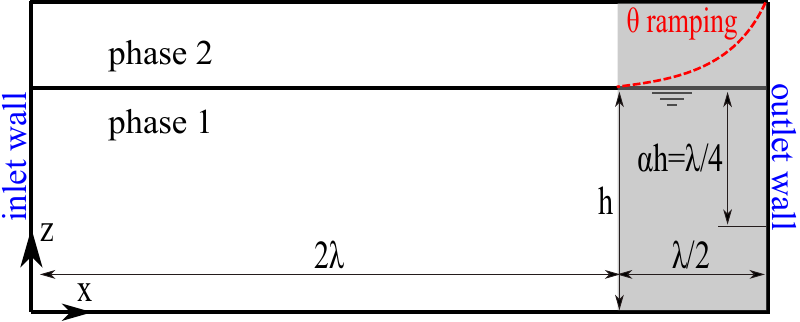}
      \caption{Schematic diagram for the numerical domain setup with an added damping zone.}
      \label{fig:numDomain02}
      \end{figure} 

	 \par The second case of choice is $\alpha h = \lambda /2$ since this value corresponds to the effective depth of a linear monochromatic deep-water wave. As can be seen in Fig. \ref{fig:envelopes}, the free surface envelope shows lower variations compared to the previous case indicating lower reflection off the absorbing wall. The reflection coefficient for this case is found to be $\epsilon_{r}=22.51\%$. This is mainly attributed to the improved matching between the absorbing velocity profile posed on the absorbing wall and the deep-water wave velocity profile; compared to the previous case.
    \par After that, the value of $\alpha h = \lambda /4$ which is predicted by the theoretical solution of Eq. (\ref{eqn:optimumAlpha}) to provide optimum absorption. Indeed, one can see in Fig. \ref{fig:envelopes} that this case provides the least variation in the wave height along the wave flume. Hence, the best absorbing wall with reflection coefficient $\epsilon_{r}=12.66\%$, which is more than $50\%$ reduction in reflection compared to the conventional shallow-water approximation setup of $\alpha h = h$. For benchmarking purposes, it is worth highlighting here that the reflection coefficient value ranged between $3.2\%$ and $11.2\%$ in the shallow-water regime using the conventional setup in \citep{pablo1}; measured using the three gauges method \citep{mansard}. 
	 \par Last case considered is when $\alpha h = \lambda /18$ which seems to be interesting for two reasons. First, it investigates the scenario when absorption depth is set to be \emph{lower} than the optimum value predicted by Eq. (\ref{eqn:optimumAlpha}). Second, it is the solution of a modified version of Eq. (\ref{eqn:optimumAlpha}) where the wave number in the left hand side is substituted by:
	\begin{equation}
	  k=\frac{\omega}{C}=\frac{\omega}{\sqrt{g \alpha h}}
	\end{equation}	  
	Where $C$ and $g$ are the wave celerity and the gravitational acceleration, respectively. This is done to take into consideration the method used in the present numerical packages, such as \texttt{IHFOAM and OlaFOAM/OlaFlow}, to find the wave celerity; as addressed earlier in the theoretical analysis section. Substituting into Eq. (\ref{eqn:optimumAlpha}) we get:
	\begin{equation}
	\sqrt{\alpha k h \cdot \tanh(kh)} =  \frac{4 \sinh(kh) [\sinh(kh) - \sinh(kh (1-\alpha)) ] }{ \sinh(2kh) + 2kh } \label{eqn:optimumAlpha02}
	\end{equation}
	Where solution to the Eq. (\ref{eqn:optimumAlpha02}) is $\alpha h \approx \lambda /18$, by trial and error, for the wave conditions considered in the present study. One can see in Fig. \ref{fig:envelopes} that wave height variation has increased compared to the previous case; the wave amplitude reflection coefficient is found to be $\epsilon_{r}=21.94\%$. Figure \ref{fig:reflectionCoe} shows the behaviour of $\epsilon_{r}$ versus $\alpha h$ for of all cases tested in the present work.
\begin{figure}[!b] 
   \centering
   \includegraphics[width=0.55\textwidth]{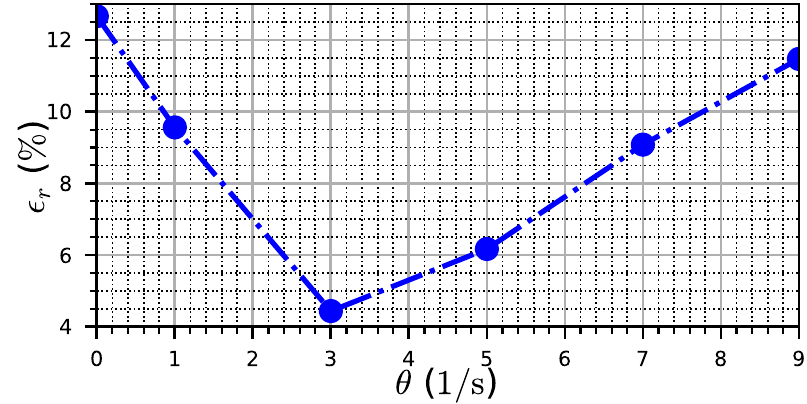} 
   \caption{Reflection coefficient ($\epsilon_{r}$) versus damping coefficient ($\theta$).}
   \label{fig:reflectionCoeWithDamping}
   \end{figure} 
\begin{figure}[!t] 
   \centering
   \includegraphics[width=\textwidth]{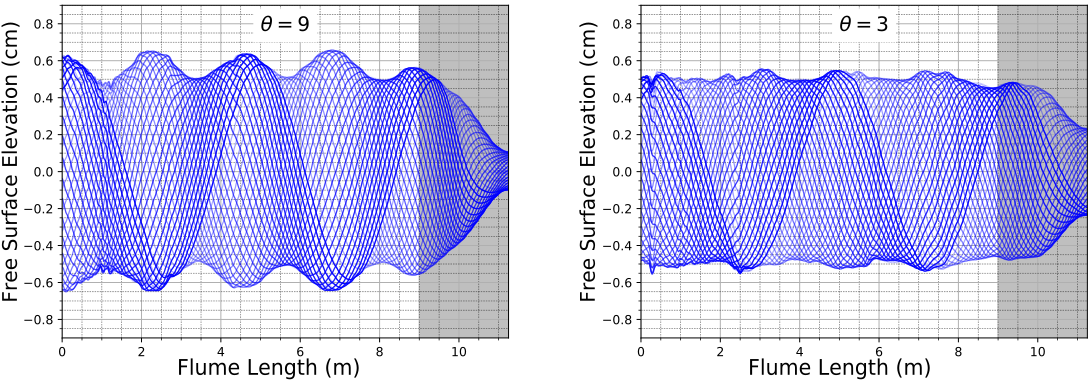}
   \caption{Temporal evolution of the water-air interface during a $10$ seconds duration for two values of the damping coefficient $\theta$.}
   \label{fig:envelopesWithDamping}
   \end{figure} 
	\par Finally, a wave gauge is installed mid-flume, i.e. at $x=4.5~\operatorname{m}$, and Fig. \ref{fig:spectrum} shows the amplitude spectra for all cases. Moreover, table \ref{tab:reflection} shows the corresponding values of the spectral peaks $A_{k}$. The table also shows the variance $\sigma^{2}$ and the significant wave height $H_{S}$ of the wave gauge reading; defined as follows:
		\begin{align}
		\sigma &= \sqrt{\frac{1}{N-1} \sum_{n=1}^{N} \zeta^{2}_{n} } \\ 
		H_{S} &= 4 \sigma
		\end{align}
		Where, $\zeta_{n}$ is the sampled water surface elevation. Additionally, the table also shows the values of the reflection coefficient $\epsilon_{r}$ using Eq. (\ref{eqn:refCoe}).
   \subsection{Absorption performance evaluation}
	\par In the preceding sections, active wave absorption was significantly improved following a hydrodynamical approach. Even though wave reflection has been dropped significantly compared to the standard shallow-water approximation setup, wave reflection is still relatively high and will adversely affect test subjects placed in that flume. The remaining reflection can be attributed to a number of reasons. First, the inherent nonlinear nature of the problem which is resembled here by the use of a fully nonlinear phase-resolving CFD model; in comparison to the linearized potential flow approach followed to derive Eq. (\ref{eqn:optimumAlpha}) and the assumption of monochromatic linear waves in the present analysis. Indeed, one can see in Fig. \ref{fig:spectrum} that waves have modulated over the spectrum by the existence of wave components beside the monochromatic fundamental frequency. 
	\par Second reason is the existing mismatch in the particle kinematic between the velocity profile of the incident waves and the one posed on the absorbing wall. For instance, the simulated waves in the present study are of an exponential velocity profile (deep-water waves) which are absorbed by a step profile on the absorbing wall Eq. (\ref{eqn:stepCurrents}). 
\par Third, since wave reflection will inevitably take place in practical applications, the value of $H$ substituted in the free surface expression $\eta(x,t)$ used in the previous sections needs to be corrected for the reflected components. In other words, take Eq. (\ref{eqn:IhOlaFoamVersion}) for instance, the subsurface velocity profile at the wall should be calculated based on the incident component only rather than the combined incident and reflected ones. For this sake, the method proposed by \citep{Ursell} to filter out the incident wave height in mechanical wavemaker setups is proposed as a remedy. First, the wave height reflection coefficient $\epsilon_{r}$ is found from measurements as addressed earlier using Eq. (\ref{eqn:refCoe}). After that, the incident wave height is set to be $H_{i}=2A$ where:
	\begin{gather}
	A = \frac{0.5 H_{\text{avg}}}{ 1 + \epsilon_{r} \cos(\delta)} 
	\end{gather}
	\begin{gather}
	\text{If $\epsilon_{r}$ is relatively small } \rightarrow A \approx 0.5 H_{\text{avg}}
	\end{gather}
	Where: $H_{\text{avg}}$ is the average wave height over half wave length of the incident wave and $\delta$ is the phase shift between the incident and reflected components. Moreover, most of the present active wave absorption numerical packages seem to disregard the existence of the evanescent modes at the vicinity of the wall, addressed earlier in Eq. (\ref{eqn:phi}), where $\eta(x,t)$ is measured on the absorbing wall itself. Experimentally, this is avoided by monitoring $\eta(x,t)$ away from the wall where evanescent modes would have diminished.
	\par Finally, the fourth major reason is the calculation of the incident waves celerity addressed earlier in the theoretical analysis section. For instance, the numerical model implemented in the present study is based on shallow-water approximation and the celerity is found correspondingly. It is worth highlighting here that this issue might be resolved, as shown in \citep{pabloX}, by solving the dispersion relation iteratively.
\begin{figure}[!t] 
   \centering
   \includegraphics[width=0.55\textwidth]{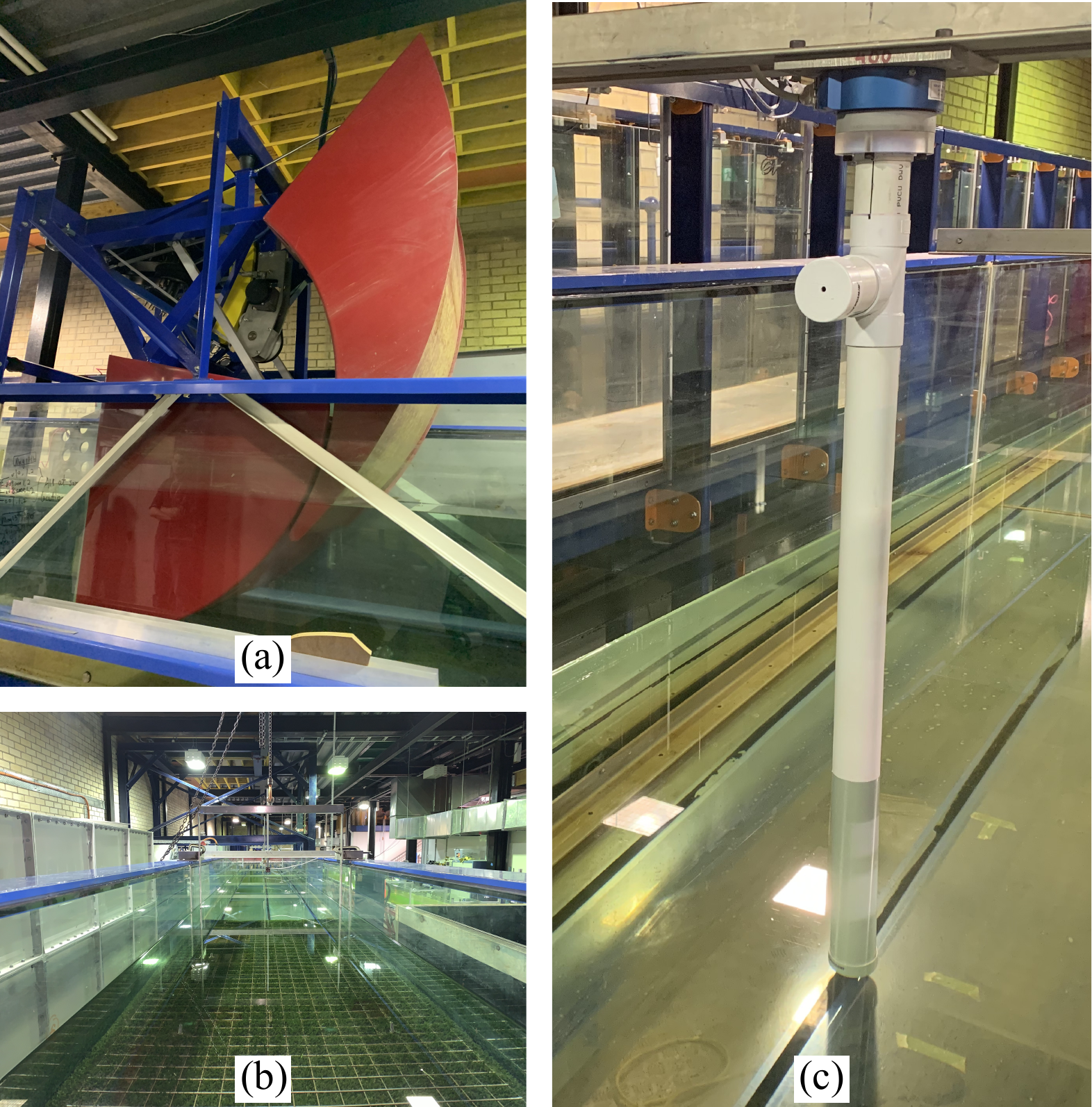}
   \caption{Experimental setup photographs for: an Edinburgh Designs wavemaker (a), a wave-absorbing beach with artificial grass (b), and a vertical rigid cylinder fitted with a force transducer (c).}
   \label{fig:expSetup}
   \end{figure} 
\begin{table}[!b]  
   \centering
   \caption{Wave-structure interaction computational domain boundary conditions in OpenFOAM's conventions.}
   \label{tab:bounCon3D}
   \resizebox{\textwidth}{!}{%
   \begin{tabular}{ccccc}
   \hline
   \multirow{2}{*}{Boundary} &       & \multicolumn{3}{c}{Boundary Condition} \\ \cline{2-5} 
                             & Field & $p$          & $U$         & $\gamma $         \\ \hline
   Inlet wall               & & \texttt{fixedFluxPressure}         & \texttt{waveVelocity}         & \texttt{waveAlpha}             \\
   top wall                 & & \texttt{totalPressure}          & \texttt{pressureInletOutletVelocity}         & \texttt{inletOutlet}             \\
   Bottom wall              & & \texttt{fixedFluxPressure}          & \texttt{slip}         & \texttt{zeroGradient}             \\
   Mid-Flume              & & \texttt{fixedFluxPressure}          & \texttt{slip}         & \texttt{zeroGradient}             \\
   Cylinder wall              & & \texttt{fixedFluxPressure}          & \texttt{slip}         & \texttt{zeroGradient}             \\
   Outlet wall $\alpha h$   & & \texttt{fixedFluxPressure}          & \texttt{waveAbsorption2DVelocity}         & \texttt{zeroGradient}             \\
   Outlet wall $(1-\alpha)h$& & \texttt{fixedFluxPressure}          & \texttt{fixedValue}         & \texttt{zeroGradient}             \\ \hline
   \end{tabular}
   }
\end{table}  
\begin{figure}[!t] 
   \centering
   \includegraphics[width=\textwidth]{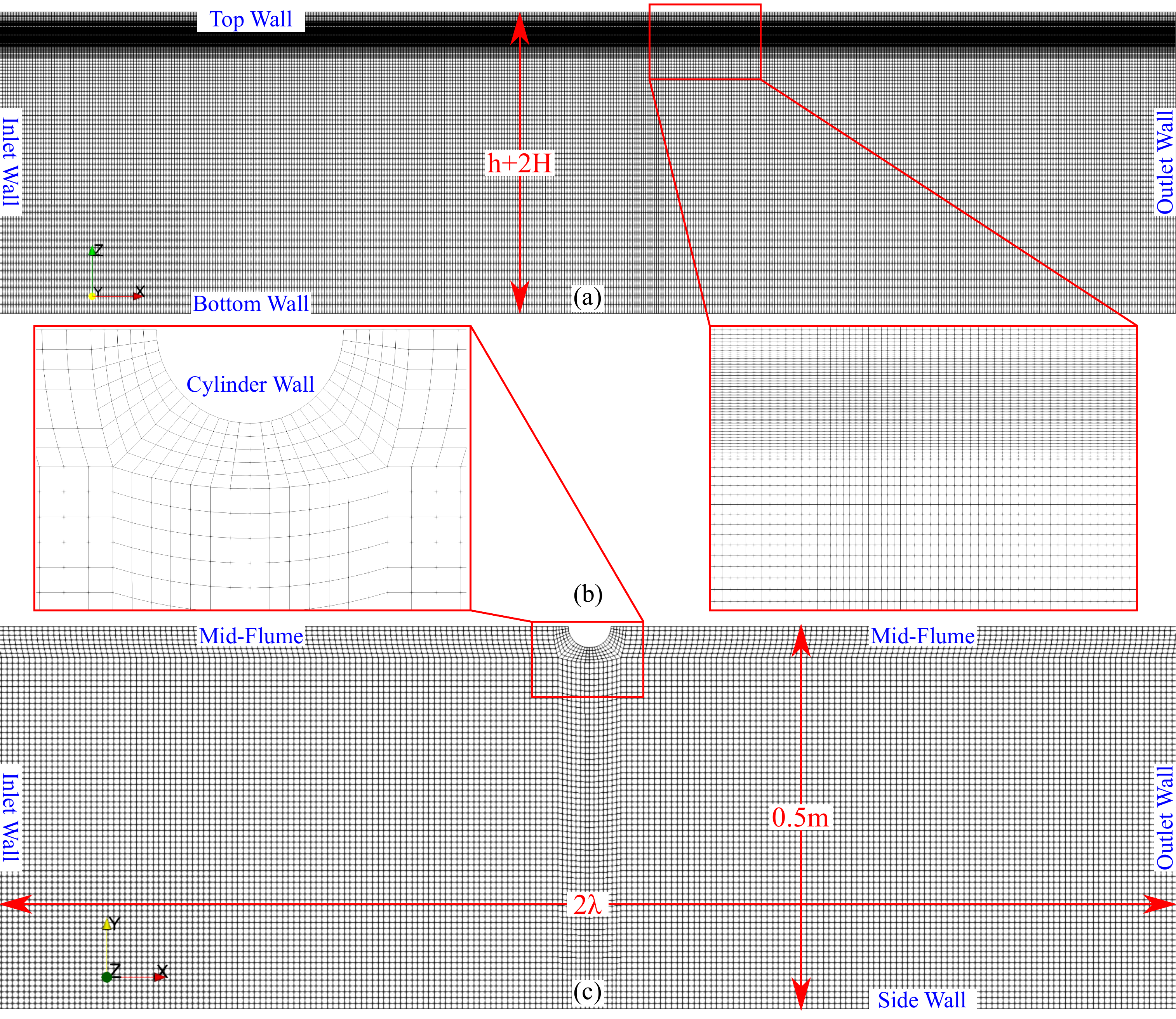}
   \caption{Snap shots for the three dimensional numerical domain showing: a vertical $x-z$ section (a), a horizontal $x-y$ section (c), and close-up views (b).}
   \label{fig:numDomain3d}
   \end{figure} 
\begin{figure}[!t] 
   \centering
   \includegraphics[width=0.85\textwidth]{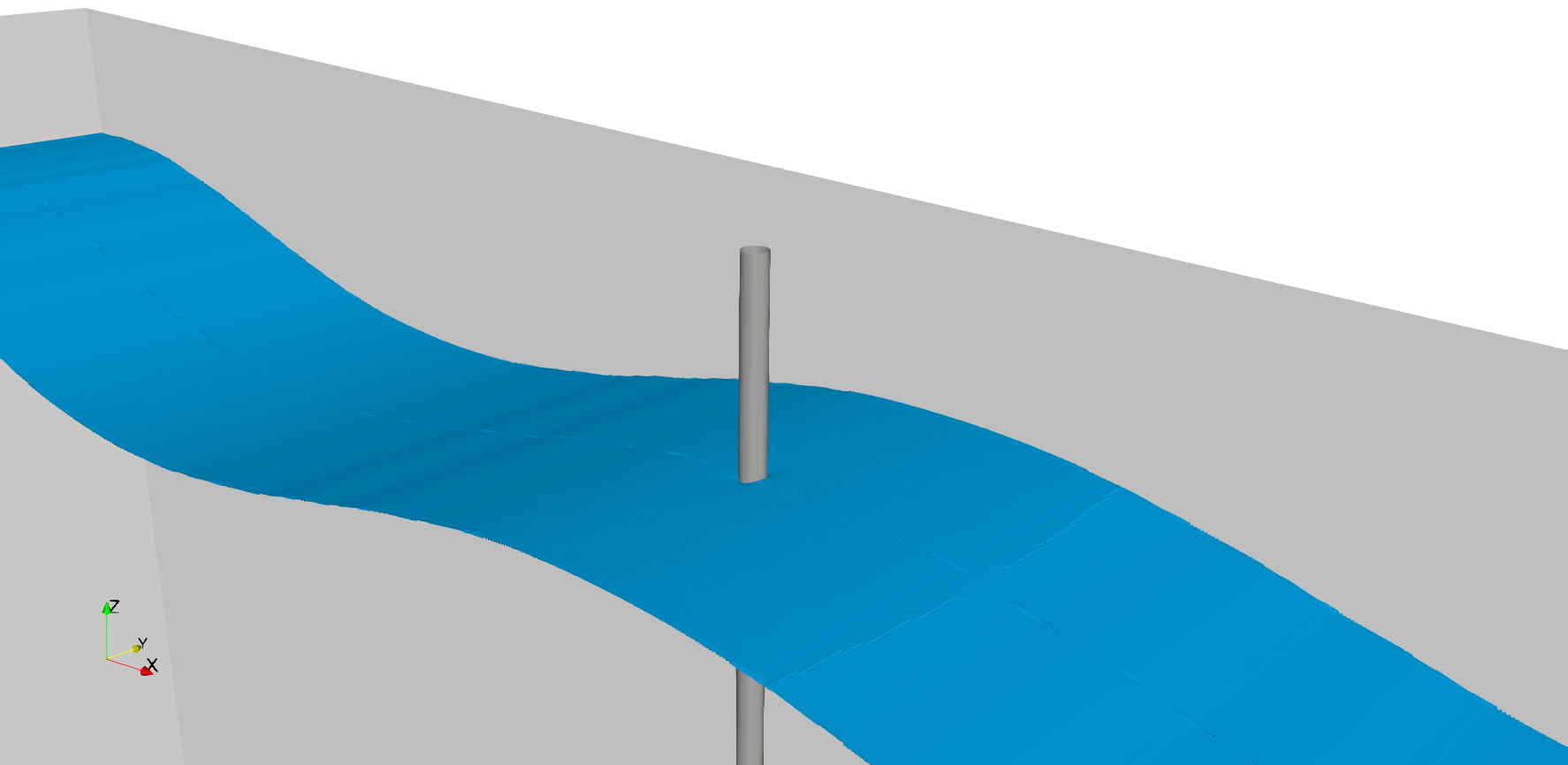}
   \caption{ A snap shot of the simulated wave passing by the vertical cylinder, using the CFD numerical model (a rendered mirror view). Free surface is represented by clipping the value of $\gamma = 0.5$.}
   \label{fig:numRes3d}
   \end{figure} 
\begin{figure}[!b] 
   \centering
   \includegraphics[width=\textwidth]{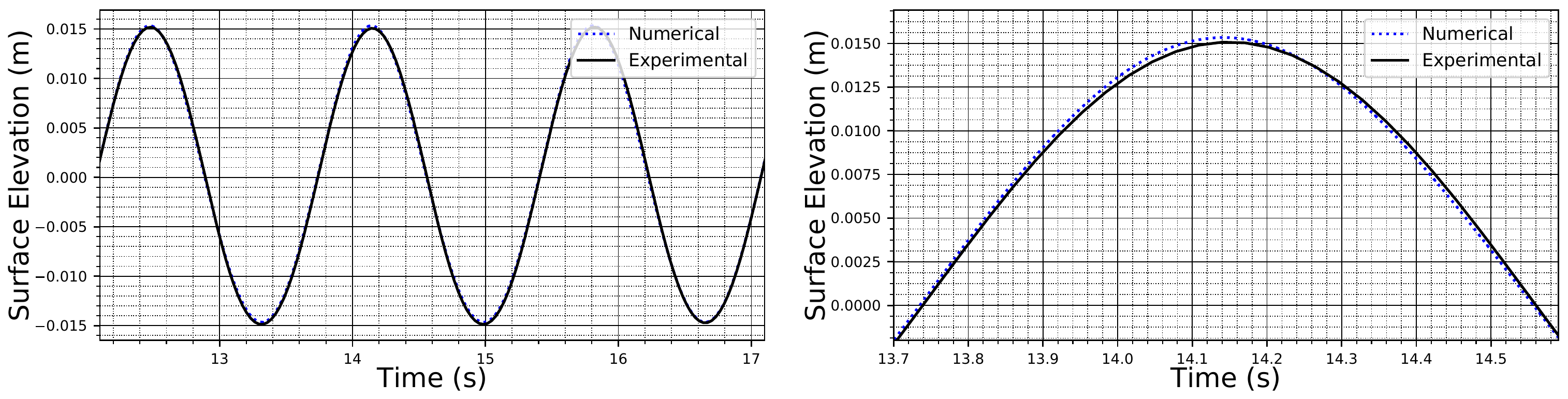}
   \caption{Numerical versus experimental free surface elevation, measured at a point placed $\lambda/2$ upstream the cylinder (a), and a close-up view (b).}
   \label{fig:wg3d}
   \end{figure} 
\begin{figure}[!t] 
   \centering
   \includegraphics[width=\textwidth]{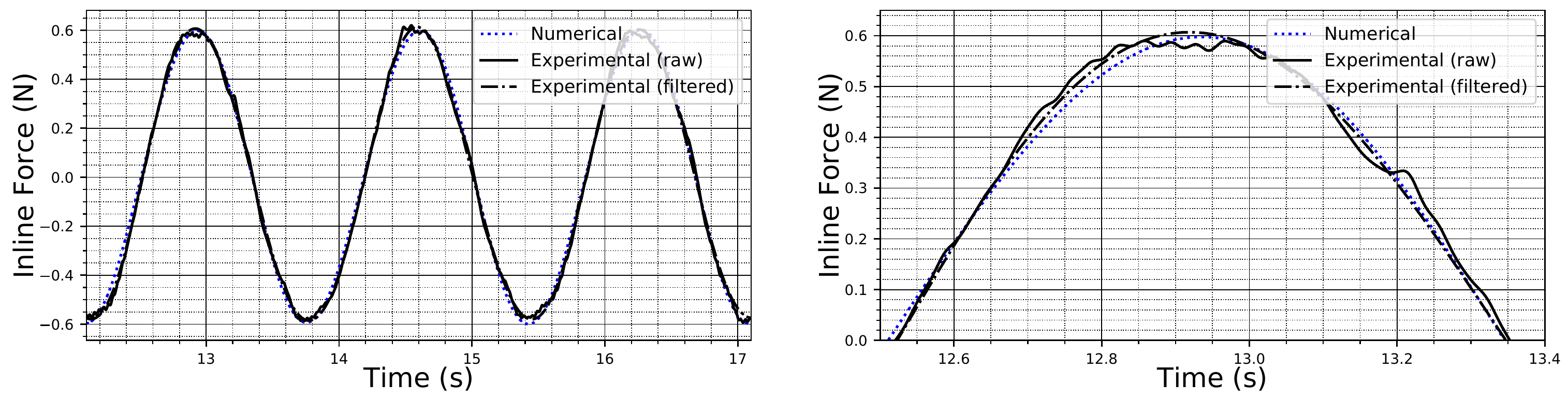}
   \caption{Numerical versus experimental inline forces on the tested cylinder (a), and a close-up view (b).}
   \label{fig:f3d}
   \end{figure} 
   \par However, even though the reflection coefficient has been dropped significantly compared to the shallow-water approximation conventional ($\alpha h = h$) setup; the remainder reflection ought to be further reduced to avoid inversely affecting tested subjects to be placed in such a flume. Remainder reflection can be reduced by adding a relatively small passive absorbing zone to amend performance of the absorbing wall; resulting in a hybrid (i.e. a combination of active and passive) approach. The underlying concept in such an approach is that the static wall is intended to absorb the majority of incident waves and the passive zone to absorb (or dissipate) the remainder wave components that have been reflected of the static absorbing wall. This simply enables the utilization of the best-of-both-worlds, where the numerical cost is significantly reduced in comparison to pure passive absorption and the reflection coefficient is being within acceptable limits in comparison to pure active absorption in deep water conditions. Other instantiations of the use of hybrid absorption approaches have been showcased in \citep{review,beach05,pabloX,hybrid02}.
   \par Consequently, a relatively small damping \emph{zone} of length $\lambda/2$ with a gradually ramped damping coefficient is added next to the $\alpha h = \lambda /4 $ absorbing wall; as shown in Fig. \ref{fig:numDomain02}. This is done by adding an artificial gradual vertical damping term to the momentum equation Eq. (\ref{eqn:ME}) such that:  
   \begin{equation}
      \frac{\partial (\rho U)}{\partial t} + \nabla \cdot (\rho U U) = - \nabla p + \rho f_{b} - \rho \theta U_{z} \label{eqn:ME_D} \\
   \end{equation}
   Where $\theta$ is the damping coefficient and $U_{z}$ is the vertical velocity component. A longer damping zone would result in less reflection but with the considerable increase in the computational cost. In fact, in pure passive absorption scenarios, it was recommended that the absorption zone should be at least $\lambda$ to $2\lambda$ in length for acceptable wave absorption \citep{review,beach05,relaxationZone6}. Consequently, the shorter damping zone was investigated in the present work to investigate the effectiveness the implemented combination. On the other hand, however, the damping coefficient seems to have an optimum value where relatively high damping coefficient $\theta$ values would increase reflection as reported in \citep{beach05,dampingTheta}.
   \par Figure \ref{fig:reflectionCoeWithDamping} shows the behaviour of the reflection coefficient ($\epsilon_{r}$) versus the damping coefficient $\theta$. The figure shows that the optimum $\theta$ value is 3 where the reflection coefficient was reduced from $12.66\%$ to $4.44\%$. However, raising the damping coefficient over that resulted in adverse absorption performance. Moreover, Fig. \ref{fig:envelopesWithDamping} shows the temporal locus of the air-water interface for two values of the damping coefficient $\theta$.

\section{Application to wave-structure interaction}
In what follows, the methodology described in the aforementioned sections is implemented in a typical engineering application; wave forcing of a rigid fixed vertical cylinder.

   \subsection{Experimental setup}
   An experimental setup was constructed in the Wave Flume at the Fluids Laboratory at the University of Sydney, Australia. The flume is $30~\operatorname{m}$ long, $1~\operatorname{m}$ wide and $1~\operatorname{m}$ deep; with a maximum water depth of $0.75~\operatorname{m}$. The experimental setup, depicted in Fig. \ref{fig:expSetup}, entails: an Edinburgh Designs piston-type wavemaker before power-up (a), a sloped wave-absorbing beach with artificial grass at the opposite end (b), and a vertical rigid cylinder fitted with a force transducer (c). The cylinder's height and diameter are $1.1~\operatorname{m}$ and $56~\operatorname{mm}$, respectively. The force transducer is a multi-axis six degrees of freedom load cell that measures forces and moments in the x, y and z directions. Finally, a twin-wire resistive wave gauge is installed about $\lambda/2$ upstream of tested cylinder.

   \subsection{Numerical model}
   A three-dimensional numerical model is constructed to model the interaction of regular wave train with a fixed vertical cylinder. The implemented governing equations and discretizations schemes are the same as those illustrated earlier in section \ref{sec:numGovEqn}. On the other hand, the numerical domain is different as shown in Fig. \ref{fig:numDomain3d}. As shown in the figure, flow symmetry is assumed around the \enquote{Mid-Flume} boundary and, therefore, half of the domain is simulated. Moreover, a multi-block structured mesh is implemented at the vicinity of the cylinder ending up with a high quality mesh as shown in the figure. The numerical domain is descretized in such a way that $500 \text{ cells}/\lambda$ in both $x$ and $y$ directions, and a maximum refinement of $21 \text{ cells}/H$ in the $z$ direction at the free surface to ensure acceptable resolution and reduce numerical diffusion \citep{beach01,theWaveLoadsProject,domDecomp}. This resulted is a total cell count of about $7.8$ million cells. Table \ref{tab:bounCon3D} summarizes the boundary conditions of choice, using OpenFOAM's conventions.

   \subsection{Test conditions and results validation}
   Different wave conditions were considered in this section, to those of section \ref{sec:GeomAndBC}, to investigate the validity of proposed approach at a different scenario. The generated waves are $0.03~\operatorname{m}$ high, $1.67~\operatorname{s}$ period, $3.7~\operatorname{m}$ long and $0.75~\operatorname{m}$ deep. Clearly, this corresponds to a steepness ($ak$) of $0.026$ and an Ursell number ($U_{r}$) of $0.99$, which indicates that the waves are slightly steeper than the proposed linear theory range of validity in \citep{mehaute}. Moreover, the relative depth of $kh=1.27$ which corresponds to the intermediate water condition. Furthermore, as seen in table \ref{tab:bounCon3D}, slip boundary conditions have been implemented which indicate that viscous effects and boundary layer formation are suppressed. This is still a valid approximation here since the simulated wave conditions here corresponds to a Keulegan–Carpenter number ($KC$) of $1.04$, which corresponds to the inertia-dominated loading regime \citep{sumer}. Substituting into Eq. (\ref{eqn:optimumAlpha}), the proposed optimum limited absorption ratio will be $\alpha=0.83$. Temporal resolution was varied in such a way that the maximum CFL number is limited to $0.15$. Finally, the simulation was run for $10.5T$. 

   \par Figure \ref{fig:numRes3d} shows the wave-structure interaction from the three-dimensional CFD model. Referring back to phase-fraction transport equation Eq. (\ref{eqn:phase}), the free surface is defined by means of clipping numerical cells where the phase-fraction $\gamma=0.5$. This illustrates the importance of opting for a high resolution mesh at the air-water interface so that we end up with relatively a sharp interface between the two material phases, rather than a smeared $\gamma$ distribution. Figure \ref{fig:wg3d} shows a comparison between the numerical and experimental wave gauge measurements at a point placed $\lambda/2$ upstream of the tested cylinder. It is observed that a very good agreement between both is  achieved, indicating proper absorption of the outlet wall. Finally, Fig. \ref{fig:f3d} shows a comparison between numerical and experimental inline forces on the cylinder. Again, a very good agreement is observed between both records, specially after filtering out noise and structural vibrations \citep{theWaveLoadsProject}. This indicate proper effectiveness of the proposed approach and its validity for engineering applications.

\section{Concluding remarks}
\par The present study aimed to extend the range of applicability of the static-boundary absorption method outside the conventional shallow-water waves limit, providing a computationally cost effective alternative to the other available methods. To tackle this, absorption of unidirectional monochromatic waves in a semi-infinite flume by means of a static wall was investigated theoretically and numerically. This was done by proposing a limited absorption depth $\alpha h$ which corresponds to the incident wave conditions, to better match the wave kinematics in deeper water conditions compared to the conventional method of the shallow-water approximated piston wavemaker. 

\par A theoretical analysis was derived for the interaction of a monochromatic wave with an absorbing wall using the wavemaker theory. The optimum absorption depth $\alpha h$ was introduced and linked to the incident wave conditions using the kinematic boundary condition on the absorbing wall; as shown in Fig. \ref{fig:optimumAlpha} and Eq. (\ref{eqn:optimumAlpha}).

\par Moreover, a nonlinear wave phase-resolving CFD numerical model was implemented to validate the proposed theoretical outcomes; using different case scenarios of the limited absorption depth $\alpha h$. As can be seen in Fig. \ref{fig:envelopes}, wave height variation along the flume was greatly influenced by the variation of the limited absorption depth $\alpha h$. Compared to the conventional shallow-water approximation setting ($\alpha h = h$), absorption was significantly enhanced where wave reflection coefficient dropped from $\epsilon_{r}=28.1\%$ to $12.66\%$. Furthermore, the proposed solution is relatively easy and straight-forward to be implemented to existing numerical packages without code modifications. In addition, it can be validated experimentally because the proposed design is, more or less, a variant of the conventional piston-type wavemaker.

\par However, even though the reflection was significantly reduced, still it should be reduced further more to avoid adversely influencing tested subjects in the CFD numerical wave flume. This remainder reflection was attributed to a number of reasons summarized in the following points: 
\begin{itemize}
\item The use of a linearized theoretical modelling to address the inherent nonlinear nature of water waves.
\item The mismatch between the incident waves and wall absorbing profiles.
\item The inclusion of both the reflected waves and evanescent modes in the measured free surface elevation $\eta(x,t)$.
\end{itemize}

\par Aside of the aforementioned caveats/limitations of this method, the remainder reflection could be further significantly reduced by means of implementing a relatively small damping zone resulting in a hybrid approach. This has resulted in reducing wave reflection form $12.66\%$ to $4.44\%$, which is comparable with typical physical model tests. 

\par Finally, the proposed corrector has been applied and experimentally validated to a typical engineering application: wave forcing of a vertical rigid cylinder. As shown in figures \ref{fig:wg3d} and \ref{fig:f3d}, a very good agreement between the numerical and experimental measurements was observed in both the free surface profile and the inline forces record.

\newglossaryentry{gls:0}{ name={$\rho$}, description={density} }
\newglossaryentry{gls:1}{ name={$U$}, description={velocity vector field variable} }
\newglossaryentry{gls:2}{ name={$ N_{i} $}, description={number of elements for the i-th grid} }
\newglossaryentry{gls:3}{ name={$ \varphi_{i} $}, description={solution variable for the i-th grid} }
\newglossaryentry{gls:4}{ name={$ r_{ij} $}, description={grid refinement ratio $N_{j}/N_{i}$} }
\newglossaryentry{gls:5}{ name={$ p $}, description={pressure field variable} }
\newglossaryentry{gls:6}{ name={$ P $}, description={apparent order of convergence} }
\newglossaryentry{gls:7}{ name={$ \varphi^{21}_{ext} $}, description={extrapolated variable} }
\newglossaryentry{gls:8}{ name={$ \phi$}, description={velocity-potential function} }
\newglossaryentry{gls:9}{ name={$ \psi $}, description={arbitrary numerical field quantity} }
\newglossaryentry{gls:10}{ name={$ e^{21}_{a} $}, description={approximate relative error} }
\newglossaryentry{gls:11}{ name={$ e^{21}_{ext} $}, description={extrapolated relative error} }
\newglossaryentry{gls:12}{ name={$ GCI^{21}_{fine} $}, description={fine-grid convergence index} }
\newglossaryentry{gls:13}{ name={$ GCI $}, description={grid convergence index} }
\newglossaryentry{gls:14}{ name={$ H_{r,i,s} $}, description={reflected, incident and significant wave heights, respectively} }
\newglossaryentry{gls:15}{ name={$ \epsilon_{r} $}, description={wave amplitude reflection coefficient} }
\newglossaryentry{gls:16}{ name={$ C $}, description={wave celerity} }
\newglossaryentry{gls:17}{ name={$ \omega $}, description={angular frequency} }
\newglossaryentry{gls:18}{ name={$ \hat{n} $}, description={unit normal vector} }
\newglossaryentry{gls:19}{ name={$ u(z,t) $}, description={horizontal velocity profile} }
\newglossaryentry{gls:20}{ name={$ S(z) $}, description={wall surface profile} }
\newglossaryentry{gls:21}{ name={$ S $}, description={stroke} }
\newglossaryentry{gls:22}{ name={$ H $}, description={wave height} }
\newglossaryentry{gls:23}{ name={$ h $}, description={water depth} }
\newglossaryentry{gls:24}{ name={$ \alpha $}, description={dimensionless absorption depth} }
\newglossaryentry{gls:25}{ name={$ \gamma $}, description={phase fraction indicator} }
\newglossaryentry{gls:26}{ name={$ k $}, description={wave number} }
\newglossaryentry{gls:27}{ name={$ f_{b} $}, description={body forces} }
\newglossaryentry{gls:28}{ name={$ A_{k} $}, description={peak spectral amplitude} }
\newglossaryentry{gls:29}{ name={$ \lambda $}, description={wave length} }
\newglossaryentry{gls:30}{ name={$ \sigma $}, description={standard deviation} }
\newglossaryentry{gls:31}{ name={$ \theta $}, description={damping coefficient} }
\glossarystyle{long} 
\glsaddall  
\printglossary[title=Nomenclature]
\bibliographystyle{unsrtnat} 
\bibliography{references}

\end{document}